\documentclass[12pt]{article}

\usepackage{latexsym}
\usepackage{amssymb,amsfonts,amsmath}
\usepackage{graphicx} 
\usepackage{indentfirst}
\usepackage{bbm}
\usepackage{amssymb}
\usepackage{verbatim}
\usepackage{amsmath, amsthm,amssymb}
\usepackage{mathrsfs}
\usepackage{hyperref}
\usepackage{amsfonts}
\usepackage{dsfont}
\usepackage{cite}
\usepackage{xcolor}
\usepackage[multiple]{footmisc}

\parskip=\medskipamount

\newcommand {\cD}{{\cal D}}
\newcommand {\cE}{{\cal E}}

\newcommand {\cJ}{{\cal J}}

\newcommand {\cM}{{\cal M}}
\newcommand {\cN}{{\cal N}}
\newcommand {\cO}{{\cal O}}

\newcommand {\cU}{{\cal U}}

\newcommand {\cW}{{\cal W}}

\def\a{\alpha}

\def\b{\beta}
\def\c{\chi}
\def\d{\delta}
\def\e{\epsilon}

\def\g{\gamma}
\def\G{\Gamma}

\def\j{\psi}
\def\k{\kappa}
\def\l{\lambda}
\def\m{\mu}
\def\n{\nu}
\def\o{\omega}
\def\p{\pi}
\def\q{\theta}
\def\r{\rho}
\def\s{\sigma}
\def\t{\tau}

\def\D{\Delta}
\def\F{\Phi}
\def\J{\Psi}
\def\L{\Lambda}
\def\O{\Omega}
\def\P{\Pi}

\def\S{\Sigma}
\def\U{\Upsilon}
\def\X{\Xi}
\def\tr{{\rm tr}}
\def\rd{{\rm d}}
\def\ri{{\rm i}}
\def\re{{\rm e}}

\def\N{{\cal N}}

\newcommand{\ad}{{\dot{\alpha}}}                           %
\newcommand{\bd}{{\dot{\beta}}}                            %
\newcommand{\ve}{\varepsilon}                            %
\newcommand{\cDB}{{\bar\cD}}                            %

\newcommand{\ab}{{\a\b}}

\renewcommand{\aa}{{\a\ad}}
\newcommand{\bb}{{\b\bd}}
\newcommand{\pa}{\partial}                           %
\newcommand{\hf}{\frac12}

\newcommand{\be}{\begin{equation}}
\newcommand{\ee}{\end{equation}}
\newcommand{\bea}{\begin{eqnarray}}
\newcommand{\eea}{\end{eqnarray}}
\newcommand{\non}{\nonumber}

\def\dt#1{{\buildrel {\hbox{\LARGE .}} \over {#1}}}    %

\newcommand{\bm}[1]{\mbox{\boldmath$#1$}}

\def\double #1{#1{\hbox{\kern-2pt $#1$}}}

\newcommand{\hm}{{\hat{m}}}

\newcommand{\ha}{{\hat{a}}}
\newcommand{\hb}{{\hat{b}}}

\newcommand{\gd}{{\dot\g}}
\newcommand{\dd}{{\dot\d}}

\newcommand{\ts}{{\tilde{\s}}}

\newif\ifdtup

\newcommand{\bsubeq}{\begin{subequations}}
\newcommand{\esubeq}{\end{subequations}}

\newcommand{\eol}{\notag \\}

\numberwithin{equation}{section}

\newcommand{\sSp}{\mathsf{Sp}}
\newcommand{\sSU}{\mathsf{SU}}
\newcommand{\sSL}{\mathsf{SL}}

\newcommand{\sSO}{\mathsf{SO}}
\newcommand{\sO}{\mathsf{O}}
\newcommand{\sU}{\mathsf{U}}

\newcommand{\sOSp}{\mathsf{OSp}}

\newcommand{\sIO}{\mathsf{IO}}

\newcommand{\id}{\mathds{1}}

\newcommand{\dalpha}{{\dot{\alpha}}}
\newcommand{\dbeta}{{\dot{\beta}}}

\newcommand{\dmu}{{\dot{\mu}}}
\newcommand{\dnu}{{\dot{\nu}}}

\newcommand{\T}{\text{T}}

\begin{document}
	
\begin{titlepage}
	\begin{flushright}
	December, 2024
	\end{flushright}
	\vspace{5mm}
	
	\begin{center}
		{\Large \bf 
			The anti-de Sitter supergeometry revisited
		}
	\end{center}

\begin{center}
		{\bf Nowar E. Koning, Sergei M. Kuzenko and Emmanouil S. N. Raptakis} \\
		\vspace{5mm}
		
		\footnotesize{ 
			{\it Department of Physics M013, The University of Western Australia\\
				35 Stirling Highway, Perth W.A. 6009, Australia}}  
		~\\
		\vspace{2mm}
		~\\
		Email: \texttt{nowar.koning@research.uwa.edu.au,
			sergei.kuzenko@uwa.edu.au, emmanouil.raptakis@uwa.edu.au }\\
		\vspace{2mm}
	\end{center}
	
\begin{abstract}
		\baselineskip=14pt
In a supergravity framework, the $\cal N$-extended anti-de Sitter (AdS) superspace in four spacetime dimensions, $\text{AdS}^{4|4\cal N} $,
 is a maximally symmetric background that is described by a curved superspace geometry 
 with structure group $\mathsf{SL}(2, \mathbb{C}) \times \mathsf{U}({\cal N})$. 
 On the other hand, within the group-theoretic setting, 
 $\text{AdS}^{4|4{\cal N}} $ is realised as the coset superspace
 $\mathsf{OSp}({\cal N}|4;\mathbb{R}) /\big[ \mathsf{SL}(2, \mathbb{C}) \times \mathsf{O}({\cal N}) \big]$, with its structure group being $\mathsf{SL}(2, \mathbb{C}) \times \mathsf{O}({\cal N})$. Here we explain how the two frameworks are related. We give two explicit realisations of 
  $\text{AdS}^{4|4{\cal N}} $ as a conformally flat superspace, thus extending the ${\cal N}=1$ and ${\cal N}=2$ results available in the literature. As applications, we describe: (i) a two-parameter deformation of the $\text{AdS}^{4|4{\cal N}} $ interval and the corresponding superparticle model;
  (ii) some implications of conformal flatness for superconformal higher-spin multiplets and an effective action generating the $\mathcal{N}=2$ super-Weyl anomaly; and (iii) $\kappa$-symmetry of the massless AdS superparticle.
\end{abstract}
\vspace{5mm}
	
	\vfill
	
	\vfill
\end{titlepage}

\newpage
\renewcommand{\thefootnote}{\arabic{footnote}}
\setcounter{footnote}{0}

\tableofcontents{}
\vspace{1cm}
\bigskip\hrule

\allowdisplaybreaks

\section{Introduction}

Recently, we have elaborated  on the geometry of $\cal N$-extended anti-de Sitter (AdS) superspace in four spacetime dimensions, $\text{AdS}^{4|4\cal N} $, by developing its description via:
(i) the embedding formalism  \cite{KT-M, KKR}; and (ii) the  supergravity-inspired framework proposed in \cite{KKR}. The embedding approach\footnote{The embedding formalism for AdS superspaces in three and five spacetime dimensions was studied in  \cite{KT,KK}. For a pedagogical review of superembeddings see \cite{Sorokin,Bandos:2023web}.}
to $\text{AdS}^{4|4\cal N} $ is a supersymmetric extension of the 
well-known realisation of AdS$_4$ as a hypersurface in ${\mathbb R}^{3,2}$ 
\bea
 -(Z^0)^2 + (Z^1)^2 + (Z^2)^2 +(Z^{3})^2 - (Z^4)^2 = -\ell^2 ={\rm const}~.
\label{EmbeddingAdS}
\eea
It makes use of the supertwistor and bi-supertwistor formulations for ${\rm AdS}^{4|4\cal N}$ introduced in \cite{KT-M} and further developed in \cite{KKR}. Its main virtues are: (i) it provides a global realisation of $\text{AdS}^{4|4\cal N} $ that is independent of the choice of local coordinates; and (ii) the manifest transitive action 
 of the AdS isometry supergroup $\sOSp(\N|4;\mathbb{R})$ on ${\rm AdS}^{4|4\cal N}$.\footnote{The supergroup $\sOSp(\N|4;\mathbb{R})$ \cite{Kac, Freund:1975nr, Nahm:1975mu, Scheunert:1976uf, Scheunert:1976wi, Rittenberg:1977eg} has $\sSp(4) \times \sO(\cN) $ as its maximal Lie subgroup. The identity component of
 the AdS isometry group $\sO(3,2)$, 
 $\sSO_0(3,2)$, is related to $\sSp(4)$ as 
$\sSO_0(3,2) \cong \sSp(4)/  \{\pm {\mathbbm 1} \}$. }
This transitive action means that ${\rm AdS}^{4|4\cal N}$ is a homogeneous superspace, 
\begin{align}
	\label{1.1}
	\text{AdS}^{4|4\N} = \frac{\sOSp(\N|4;\mathbb{R})}{\sSL(2,\mathbb{C})\times\sO(\N)}\,,
\end{align}
which generalises the group-theoretic realisation of AdS${}_4$ as a homogeneous space of $\sO(3,2)$.

In the $\cN=1$ case, coset superspace \eqref{1.1} was introduced in the 1970s by Keck \cite{Keck} and Zumino \cite{Zumino77}, and the comprehensive analysis of general supermultiplets on $\text{AdS}^{4|4}$ was given in a paper by Ivanov 
and Sorin \cite{IS}, arguably one of the most important works on AdS 
supersymmetry.\footnote{The structure equations for $\rm{AdS}^{4|4} $ were described in \cite{IS} and more recently in \cite{BILS}, see also \cite{Bandos:1999qf, Bandos:1999pq}.}
Years ago, it was also recognised that  $\rm{AdS}^{4|4} $ originates as a maximally supersymmetric solution in the following off-shell formulations for $\cN=1$ supergravity: (i)  the old minimal supergravity \cite{Siegel77-77,WZ,old1,old2} with a cosmological term \cite{Townsend}, 
see \cite{GGRS,BK} for a review; and (ii) the non-minimal AdS supergravity \cite{BK11}.
Analogous results exist for $\cN=2$ AdS superspace \cite{KLRT-M1,KT-M08,Butter:2011ym}.

The supergravity-inspired approach to $\text{AdS}^{4|4\cN}$ advocated in \cite{KKR} is based on the concept 
of $\cN$-extended conformal superspace with flat connection \cite{Kuzenko:2021pqm}.
The virtue of this geometric setting is that it can be used to describe every conformally flat superspace. We define a curved superspace to be conformally flat if its covariant derivatives can locally be turned into those corresponding to Minkowski superspace ${\mathbb M}^{4|4\cN}$ by applying a local $\sSU(2,2|\cN)$ transformation. 
Given a conformally flat superspace, this definition will be shown to imply that a local frame may be chosen 
in which the covariant derivatives $\cD_A = (\cD_a, \cD_\a^i , \bar \cD^\ad_i) $ are related to flat superspace ones, $D_A=(\pa_a, D_\a^i , \bar D^\ad_i)$, by the rule: 
\begin{subequations}
	\label{AdSBoost}
	\begin{align}
		\cD_\a^i &= \re^{\frac{\cN-2}{2\cN} \s + \frac 1 \cN \bar{\s}} \Big( D_\a^i+ D^{\b i}\s M_{\a \b} + D_{\a}^j \s \mathbb{J}^{i}{}_j \Big) ~, \label{3.13a} \\
		\bar{\cD}_{i}^{\ad}&=\re^{\frac{1}{\cN} \s + \frac{\cN-2}{2\cN} \bar{\s}} \Big( \bar{D}^\ad_i- \bar{D}_{ \bd i} \bar{\s} \bar{M}^{\ad \bd} - \bar{D}^{\ad}_j \bar{\s} \mathbb{J}^{j}{}_i \Big)~,
			\\
		\cD_\aa &= \re^{\hf \s + \hf \bar{\s}} \Big(\partial_\aa + \frac{\rm i}{2} D^i_{\a} \s \bar{D}_{\ad i} + \frac{\rm i}{2} \bar{D}_{\ad i} \bar{\s} D_{\a}^i + \hf \Big( \partial^\b{}_\ad (\s + \bar \s ) - \frac{\ri}{2} D^{\b i} \s \bar{D}_{\ad i} \bar{\s} \Big) M_{\a \b} \non \\ & \qquad \qquad \quad + \hf \Big( \partial_{\a}{}^\bd (\s + \bar{\s}) + \frac{\ri}{2} D_{\a}^i \s \bar{D}^{\bd}_i \bar{\s} \Big) { \bar M}_{\ad \bd} \Big)~. \label{3.13c}
	\end{align}
\end{subequations}
Here $(M_{\a\b} , \bar M_{\ad \bd})$ are the Lorentz generators,  $\mathbb{J}^{i}{}_j $
the $\sSU(\cN)$ generators, and the super-Weyl parameter $\s$ is a chiral superfield, 
$\bar D^\ad_i \s =0$. 
As an application of this formalism, we demonstrate below that $\text{AdS}^{4|4\cN}$ is conformally flat, by solving the equations on $\s$ which single out the AdS supergeometry. This result agrees with  the discussion in the literature \cite{BILS} that $\text{AdS}^{4|4\cN}$ is locally conformally flat, although an explicit solution of the form \eqref{AdSBoost}  was not derived for $\cN>1$ in \cite{BILS}.

More generally, the conformal superspace approach in four dimensions \cite{ButterN=1,ButterN=2, ButterN=4, KR23} is a powerful formalism to describe off-shell $\cN$-extended supergravity-matter couplings 
with $\cN\leq4$.\footnote{The $\cN=1$ and $\cN=2$ formulations are reviewed in \cite{KRT-M_N=1} and \cite{KRT-M_N=2}, respectively.} 
Conceptually, it is a superspace counterpart of the superconformal tensor calculus, see, e.g., \cite{FVP} for a review. 
In the absence of matter and compensator multiplets, conformal superspace is the gauge theory of the superconformal group $\sSU(2,2|\cN)$ and describes conformal supergravity. The torsion and curvature tensors of conformal superspace are expressed in terms of the so-called super-Weyl tensor and its covariant derivatives.  Hence, when the super-Weyl tensor vanishes, the connection is flat. 
Beyond $\cN=4$, only the conformal superspace with a flat connection \cite{Kuzenko:2021pqm} is known.
It should be pointed out that  the conformal superspace approach 
has also been extended to two \cite{Kuzenko:2022qnb},  three \cite{BKNT-M1, BKNT-M2, KNT-M},  five \cite{ BKNT-M15} and six \cite{BKNT} dimensions. 

In order to connect the $\cN$-extended conformal superspace to the conventional formulation for conformal supergravity \cite{Howe}, one has to perform a procedure of degauging \cite{ButterN=1,ButterN=2,  KR23} 
that: (i) reduces the structure group to $\sSL(2,\mathbb{C}) \times \sU(\cN)_R$; 
and (ii) converts the local scale symmetry into  super-Weyl freedom. Even in the case of conformal superspace with flat connection, the degauging results in the appearance of non-vanishing torsion and curvature tensors. 
The AdS supergeometry is singled out by the conditions that these tensors be: (i) Lorentz invariant; and (ii) covariantly constant.

Upon degauging,  the structure group is $\sSL(2,\mathbb{C}) \times \sU(\cN)_R$ and 
the superspace covariant derivatives $\cD_A $ include two connections, the Lorentz and $\sU(\cN)$ ones. On the other hand, within the group-theoretic setting, 
 $\text{AdS}^{4|4{\cal N}} $ is realised as the coset superspace \eqref{1.1}
 with its structure group being $\mathsf{SL}(2, \mathbb{C}) \times \mathsf{O}({\cal N})_R$. Here we explain how the two frameworks are related. This issue was not analysed in \cite{KKR}.
In addition, we also present several extensions to the results of \cite{KKR}. In particular, we provide two conformally flat realisations of $\text{AdS}^{4|4\cN}$
of the form \eqref{AdSBoost}, corresponding to stereographic and Poincar\'e coordinates. 
Further, we compute the vielbein for the coset superspace \eqref{1.1} in a local coordinate chart introduced in \cite{KKR}. The explicit structure of the obtained vielbein shows that a conformally flat representation for the AdS covariant derivatives does not appear to exist when dealing with structure group $\mathsf{SL}(2, \mathbb{C}) \times \mathsf{O}({\cal N})$ corresponding to the coset \eqref{1.1}.

This paper is organised as follows. We begin in section \ref{section2} by reviewing the supergravity approach to conformally flat supergeometries developed in \cite{KKR}. This forms the foundation for our studies of $\cN$-extended AdS superspaces in section \ref{section3}.  Concluding comments and applications of the obtained results are given in section \ref{section5}. The main body of this paper is accompanied by three technical appendices. 
In appendix \ref{AppendixA} our conventions for the $\cN$-extended superconformal algebra are spelt out.
Appendix \ref{section4} demonstrates that the coset construction based on the use of \eqref{1.1} does not provide a conformally flat frame for $\cN \geq 1$.
Appendix \ref{AppendixC} discusses some implications of conformal flatness in the $\cN=2$ case.

\section{Conformally flat supergeometry}
\label{section2}

In our previous work \cite{KKR}, we described the most general conformally flat $\cN$-extended supergeometry in four dimensions, building on the earlier work \cite{Kuzenko:2021pqm} which introduced the $\cN$-extended conformal superspace with a flat connection. 
This section is devoted to a review of the salient details of this framework.

\subsection{Conformal superspace with flat connection}
\label{section2.1}

In this subsection we provide a brief review of the $\cN$-extended conformal superspace with a flat connection \cite{Kuzenko:2021pqm}.%
\footnote{The conformal superspace approach to describe $\cN \leq 3$ conformal supergravity in four dimensions was developed in  \cite{ButterN=1,ButterN=2,KR23}, and its $\cN=4$ extension has been sketched in \cite{ButterN=4}. The formulations for the $\cN=1$ and $\cN=2$ cases are reviewed in \cite{KRT-M_N=1} and \cite{KRT-M_N=2}, respectively. Beyond $\cN=4$, only the conformal superspace with a flat connection \cite{Kuzenko:2021pqm} is known.}
Our starting point will be an $\N$-extended superspace $\cM^{4|4\N}$, parametrised by local coordinates 
$z^{M} = (x^{m},\theta^{\m}_\imath,\bar \theta_{\dot{\mu}}^\imath)$, where $m=0, 1, 2, 3$, $\mu = 1, 2$, $\dot{\mu} = \dot{1}, \dot{2}$ and
$\imath = \underline{1}, \dots, \underline{\cN}$. We take the structure group to be the superconformal group $\sSU(2,2|\cN)$.
Its corresponding Lie superalgebra, $\mathfrak{su}(2,2|\cN)$, is spanned by the super-translation $P_A=(P_a, Q_\a^i ,\bar Q^\ad_i)$, Lorentz $M_{ab}$,  $R$-symmetry
$\mathbb{Y}$ and $\mathbb{J}^{i}{}_j$, dilatation $\mathbb{D}$,  and the special superconformal $K^A=(K^a, S^\a_i ,\bar S_\ad^i)$ generators.\footnote{Our conventions for $\mathfrak{su}(2,2|\cN)$ are collected in appendix \ref{AppendixA}.}
The geometry of this superspace is encoded within the conformally covariant derivatives $\nabla_A = (\nabla_a,\nabla_\a^i,\bar{\nabla}_i^\ad)$, which take the form:
\begin{align}
	\label{6.1}
	\nabla_A &= E_A{}^M \partial_M - \hf \Omega_A{}^{bc} M_{bc} - \Phi_A{}^j{}_k \mathbb{J}^{k}{}_j - \ri \Phi_A \mathbb{Y}
	- B_A \mathbb{D} - \frak{F}_{AB} K^B \eol
	&= E_A{}^M \partial_M - \Omega_A{}^{\b\g} M_{\b\g} - \bar{\Omega}_A{}^{\bd\gd} \bar{M}_{\bd\gd}
	- \Phi_A{}^j{}_k \mathbb{J}^{k}{}_j - \ri \Phi_A \mathbb{Y} - B_A \mathbb{D} - \frak{F}_{A B} K^B ~,
\end{align} 
where $E_{A}{}^M$ denotes the inverse supervielbein while the remaining superfields are connections associated with the non-translational generators of the superconformal group.

By definition, the gauge group of conformal supergravity is generated by local transformations of the form
\begin{align}
	\label{6.2}
	\nabla_A' = \re^{\mathscr{K}} \nabla_A \re^{-\mathscr{K}} ~, \qquad
	\mathscr{K} =  \xi^B \nabla_B+ \hf K^{bc} M_{bc} + \S \mathbb{D} + \ri \rho \mathbb{Y} 
	+ \chi^{i}{}_j \mathbb{J}^{j}{}_i + \L_B K^B ~ ,
\end{align}
where the gauge parameters satisfy natural reality conditions. Given a conformally covariant tensor superfield $\cU$ (with its indices suppressed), it transforms under such transformations as follows:
\begin{align}
	\label{6.3}
	\cU' = \re^{\mathscr{K}} \cU~.
\end{align}

Within the conformal superspace approach to $\cN$-extended conformal supergravity with $ \cN\leq 4$
\cite{ButterN=1,ButterN=2,KR23,ButterN=4}, the graded commutator 
$[\nabla_A,\nabla_B\}$ is expressed in terms of the corresponding super-Weyl tensor and its covariant derivatives. The super-Weyl tensor $\cW_{\a_1\dots \a_{4-\cN}}$
is covariantly chiral in the  $\cN<4$ case \cite{ButterN=1,ButterN=2,KR23}; its structure is more involved for $\cN=4$.
For $\cN$-extended conformal superspace with a flat connection, the graded commutator 
$[\nabla_A,\nabla_B\}$ takes the flat-superspace form 
 \cite{Kuzenko:2021pqm}, which means 
\bea
\label{CSSCFlat}
\{ \nabla_\a^i , \bar{\nabla}^{\bd}_j \} = - 2 \ri \d^i_j \nabla_\a{}^{\bd} ~,
\eea
and the other  (anti)commutators are equal to zero. 
Since the super-Weyl tensor vanishes, applying a gauge transformation \eqref{6.2}
allows one  (at least locally) to turn the covariant derivatives $\nabla_A$ into 
$ D_{A} = (\partial_{a}, D^i_{\a}, \bar{D}_i^{\ad})$ corresponding to $\N$-extended Minkowski superspace ${\mathbb M}^{4|4\cN}$.

\subsection{Degauging (i): $\sU(\cN)$ superspace}
\label{section2.2}

According to eq. \eqref{6.2}, under an infinitesimal special superconformal gauge transformation $\mathscr{K} = \Lambda_{B} K^{B}$, the dilatation connection transforms as follows
\bea
\d_{\mathscr{K}} B_{A} = - 2 \Lambda_{A} ~.
\eea
As a result, it is possible to impose the gauge
$B_{A} = 0$, completely fixing
the special superconformal gauge freedom.\footnote{Actually, there is a class of residual gauge transformations which preserve this gauge. They lead to the super-Weyl transformations of the degauged geometry.} Hence, the corresponding connection is no longer required for the covariance of $\nabla_A$ under the residual gauge freedom and
may be
extracted from $\nabla_{A}$,
\bea
\nabla_{A} &=& \cD_{A} - \mathfrak{F}_{AB} K^{B} ~. \label{ND}
\eea
Here the operator $\cD_{A} $ involves only the Lorentz and $R$-symmetry connections
\bea
\cD_A = E_A{}^M \partial_M - \frac{1}{2} \O_A{}^{bc} M_{bc} - \Phi_A{}^j{}_k \mathbb{J}^{k}{}_j - \ri \F_A \mathbb{Y}~.
\eea

The next step is to relate the special superconformal connection
$\mathfrak{F}_{AB}$  to the torsion tensor associated with $\cD_A$. To do this, one can  make use of the relation
\bea
\label{4.3}
[ \cD_{A} , \cD_{B} \} &=&  [ \nabla_{A} , \nabla_{B} \} + \big(\cD_{A} \mathfrak{F}_{BC} - (-1)^{\e_A \e_B} \cD_{B} \mathfrak{F}_{AC} \big) K^C + \mathfrak{F}_{AC} [ K^{C} , \nabla_B \} \non \\
&& - (-1)^{\e_A \e_B} \mathfrak{F}_{BC} [ K^{C} , \nabla_A \} - (-1)^{\e_B \e_C} \mathfrak{F}_{AC} \mathfrak{F}_{BD} [K^D , K^C \} ~.
\eea
In conjunction with the algebra \eqref{CSSCFlat}, this leads to a set of consistency conditions that are equivalent to the Bianchi identities of $\sU(\cN)$ superspace \cite{Howe} with vanishing super-Weyl tensor. Their solution expresses the components of $\mathfrak{F}_{AB}$ in terms of the torsion tensor of $\sU(\cN)$ superspace and determines the algebra $[ \cD_{A} , \cD_{B} \}$. We omit such an analysis here and instead simply present the geometry of $\cD_A$ below. The interested reader is referred to \cite{KKR} for the complete analysis.

\subsubsection{$\cN=1$ case}

In the $\cN=1$ case, the algebra of covariant derivatives \eqref{4.3} may be brought to the form\footnote{We emphasise that this algebra will not coincide with \eqref{4.3}. This is because we have simplified the geometry by performing the shift $\cD_{\a \ad} \rightarrow \cD_{\a \ad} + \frac{\ri}{2} G^{\b}{}_{\ad} M_{\a \b} - \frac{\ri}{2} G_{\a}{}^{\bd} \bar{M}_{\ad \bd} - \frac{\ri}{4} G_{\a \ad} \mathbb{Y}$.}
\begin{subequations} \label{U(1)algebra}
	\bea
	\{ \cD_{\a}, \cD_{\b} \} &=& -4{\bar R} M_{\a \b}~, \qquad
	\{\cDB_{\ad}, \cDB_{\bd} \} =  4R {\bar M}_{\ad \bd}~, \label{U(1)algebra.a}\\
	&& {} \qquad \{ \cD_{\a} , \cDB_{\ad} \} = -2{\rm i} \cD_{\a \ad} ~, 
	\label{U(1)algebra.b}	\\
	\big[ \cD_{\a} , \cD_{ \b \bd } \big]
	& = &
	{\rm i}
	{\ve}_{\a \b}
	\Big({\bar R}\,\cDB_\bd + G^\g{}_\bd \cD_\g
	- \cD^\g G^\d{}_\bd  M_{\g \d}
	\Big)
	+ {\rm i} \cDB_{\bd} {\bar R}  M_{\a \b}
	\non \\
	&&
	-\frac{\ri}{3} \ve_{\a\b} \bar X^\gd \bar M_{\gd \bd} - \frac{\ri}{6} \ve_{\a\b} \bar X_\bd \mathbb{Y}
	~, \label{U(1)algebra.c}\\
	\big[ {\bar \cD}_{\ad} , \cD_{\b\bd} \big]
	& = &
	- {\rm i}
	\ve_{\ad\bd}
	\Big({R}\,\cD_{\b} + G_\b{}^\gd \cDB_\gd
	- \cDB^{\gd} G_{\b}{}^{\dd}  \bar M_{\gd \dd}
	\Big) 
	- {\rm i} \cD_\b R  {\bar M}_{\ad \bd}
	\non \\
	&&
	+\frac{\ri}{3} \ve_{\ad \bd} X^{\g} M_{\g \b} - \frac{\ri}{6} \ve_{\ad\bd} X_\b \mathbb{Y}
	~. \label{U(1)algebra.d}
	\eea
\end{subequations}
Here $R$ is a chiral scalar superfield
\begin{subequations}
	\label{N=1BIs}
	\bea
	\bar{\cD}_{\ad} R = 0 ~, \qquad \mathbb{Y} R = -6 R~, 
	\eea
	while $X_\a$ is the chiral field strength of a $\sU(1)$ vector multiplet
	\bea
	\bar{\cD}_{\ad} X_{\a} = 0 ~, \qquad \cD^\a X_\a = \bar{\cD}_\ad \bar{X}^\ad~, \qquad \mathbb{Y} X_\a = - 3X_\a~,
	\eea
	and $G_{\a \ad}$ is a real vector superfield. These are related via
	\bea
	X_{\a} &=& \cD_{\a}R - \bar{\cD}^{\ad}G_{\a \ad} ~, \label{Bianchi1} \\
	{\rm i} \cD_{(\a}{}^{\gd} G_{\b ) \gd} &=&  \frac{1}{3} \cD_{(\a} X_{\b)} ~.
	\eea
\end{subequations}
This supergeometry is a $\sU(1)$ superspace \cite{Howe,GGRS} with vanishing super-Weyl tensor.

Above we made use of the special superconformal gauge freedom to degauge from conformal to $\sU(1)$ superspace by fixing the gauge $B_A=0$. In this gauge, one may perform certain combined special superconformal and dilatation transformations which maintain a vanishing dilatation connection. Specifically,
\begin{align}
	\mathscr{K}(\S) = \S \mathbb{D} + \hf \nabla_B \S K^{B} \quad \implies \quad B'_A = 0~,
\end{align}
where $\S = \bar{\S}$. Then, by making use of the following relation
\bea
\nabla'_{A} &=& \cD'_{A} - \mathfrak{F}'_{AB} K^{B} = \re^{\mathscr{K}(\S)} \nabla_A \re^{-\mathscr{K}(\S)}~,
\eea
we can deduce the transformations it induces on $\cD_A$ and the torsions of $\sU(1)$ superspace. The result is as follows:
\begin{subequations}\label{FinitesuperWeylTf}
	\bea
	\cD_{\a}' & = & \re^{\frac{1}{2} \S} \left( \cD_{\a} + 2 \cD^{\b} \Sigma M_{\b \a} - \frac{1}{2} \cD_{\a} \Sigma \mathbb{Y} \right) ~, \\
	\cDB_{\ad}' & = & \re^{\frac{1}{2} \S} \left( \cDB_{\ad} + 2 \cDB^{\bd} \S {\bar M}_{\bd \ad} + \frac{1}{2} \cDB_{\ad} \S \mathbb{Y} \right) ~, \\
	\cD_{\a \ad}' & = & \re^{\S} \Big( \cD_{\a \ad} + {\rm i} \cD_{\a} \S \cDB_{\ad} + {\rm i} \cDB_{\ad} \S \cD_{\a} + {\rm i} \left( \cDB_{\ad} \cD^{\b} \S  + 2 \cDB_{\ad} \S \cD^{\b} \S \right) M_{\b \a} \non \\ && + {\rm i} \left( \cD_{\a} \cDB^{\bd} \S + 2 \cD_{\a} \S \cDB^{\bd} \S \right) { \bar M}_{\bd \ad} + {\rm i} \Big( \frac{1}{4} \left[ \cD_{\a} , \cDB_{\ad} \right] \S +  \cD_{\a} \S \cDB_{\ad} \S \Big) \mathbb{Y} \Big) ~.~~~ \\
	R' & = & \re^{\S} \Big( R + \frac{1}{2} \cDB^{2} \S - ( \cDB \S )^{2} \Big) ~, \\
	G_{\a \ad}' & = &  \re^{\S} \Big( G_{\a \ad} + [ \cD_{\a} , \cDB_{\ad} ] \S + 2 \cD_{\a} \S \cDB_{\ad} \S \Big) ~, \\
	X_{\a}' & = & \re^{\frac{3}{2} \S} \Big( X_{\a} 
	- \frac{3}{2} (\cDB^{2} - 4 R) \cD_{\a} \S \Big) ~,
	\label{swx}
	\eea
\end{subequations}
which are the (finite) super-Weyl transformations of $\sU(1)$ superspace \cite{KR19}. For infinitesimal $\S$, they reduce to the infinitesimal super-Weyl transformations presented in \cite{Howe}.

\subsubsection{$\cN>1$ case}

As was shown in \cite{KKR}, it follows from eq. \eqref{4.3} that for $\cN>1$ the algebra of degauged spinor covariant derivatives takes the form:\footnote{In the $\cN=2$ case, the torsion tensor $Y_{\a \b}^{ij}$ is reducible and should be replaced with $\hf \ve^{ij} Y_{\a \b}$.}
\begin{subequations}
	\label{U(N)algebra}
	\bea
	\{ \cD_\a^i , \cD_\b^j \}
	&=&
	4 S^{ij}  M_{\a\b} 
	+4\ve_{\a\b} Y^{ij}_{\g\d}  M^{\g\d}  
	-4\ve_{\a \b} S^{k[i} \mathbb{J}^{j]}{}_k
	+ 8{Y}_{\a\b}^{k(i}  \mathbb{J}^{j)}{}_k
	~,
	\\
	\label{U(N)algebra-b}
	\{ \cD_\a^i , \bar{\cD}^\bd_j \}
	&=&
	- 2 \ri \d_j^i\cD_\a{}^\bd
	+4\Big(
	\d^i_jG^{\g\bd}
	+\ri G^{\g\bd}{}^i{}_j
	\Big) 
	M_{\a\g} 
	+4\Big(
	\d^i_jG_{\a\gd}
	+\ri G_{\a\gd}{}^i{}_j
	\Big)  
	\bar{M}^{\bd\gd}
	\non\\
	&&
	+8 G_\a{}^\bd \mathbb{J}^i{}_j
	+4\ri\d^i_j G_\a{}^\bd{}^{k}{}_l \mathbb{J}^l{}_{k}
	-2\Big(
	\d^i_jG_\a{}^\bd
	+\ri G_\a{}^\bd{}^i{}_j
	\Big)
	\mathbb{Y} 
	~.
	\eea
	\esubeq
	The dimension-1 superfields introduced above have the following symmetry properties:  
	\bea
	S^{ij}=S^{ji}~, \qquad Y_{\a\b}^{ij}=Y_{\b\a}^{ij}=-Y_{\a \b}^{ji}~, \qquad {G_{\a \ad}{}^{i}{}_i} = 0~,
	\eea
	and satisfy the reality conditions
	\bea
	\overline{S^{ij}} =  \bar{S}_{ij}~,\quad
	\overline{Y_{\a\b}^{ij}} = \bar{Y}_{\ad\bd ij}~,\quad
	\overline{G_{\b\ad}} = G_{\a\bd}~,\quad
	\overline{G_{\b\ad}{}^{i}{}_j} = - G_{\a\bd}{}^j{}_{i}
	~.~~~~~~
	\eea
	The ${\sU}(1)_R$ charges of the complex superfields are:\footnote{We note that these torsions are uncharged for $\cN=4$. This follows from $\mathbb{Y}$ acting as a central charge in this case.}
	\bea
	{\mathbb Y} S^{ij}=\frac{2(4-\cN)}{\cN}S^{ij}~,\qquad
	{\mathbb Y}  Y^{ij}_{\a\b}=\frac{2(4-\cN)}{\cN}Y^{ij}_{\a\b}~.
	\eea
	Further, they satisfy the Bianchi identities:
	\begin{subequations}\label{BI-U2}
		\bea
		\cD_{\a}^{(i}S^{jk)}&=&0~, \quad
		\bar{\cD}_{\ad i}S^{jk} - \ri\cD^{\b (j}G_{\b\ad}{}^{k)}{}_i = \frac{1}{\cN+1} \d_i^{(j} \Big( 2 \bar{\cD}_{\ad l} S^{k)l} - \ri {\cD}^{\b |l|} G_{\b \ad}{}^{k)}{}_l\Big) ~, ~~~
		\\
		\cD_{(\a}^{(i}Y_{\b\g)}^{j)k}&=&0~, \quad
		\cD^{\b k} Y_{\a \b}^{ij} = - \cD_\a^{[i} S^{j]k}~, \quad
		\bar{\cD}_j^\bd Y_{\a \b}^{ij} = 2 \cD_{(\a}^i G_{\b)}{}^{\bd} - \ri \frac{\cN-2}{\cN+1} \cD_{(\a}^j G_{\b)}{}^{\bd i}{}_j~,~~~ \\
		\cD_{(\a}^{(i}G_{\b)\bd}{}^{j)}{}_k&=&\frac{1}{\cN+1} \cD_{(\a}^l G_{\b) \bd}{}^{(i}{}_l \d^{j)}_k~, 
		\qquad
		\cD_{(\a}^{[i}G_{\b)\bd}{}^{j]}{}_k=-\frac{1}{\cN-1} \cD_{(\a}^l G_{\b) \bd}{}^{[i}{}_l \d^{j]}_k~, 
		\\
		\cD_\a^iG^{\a \bd}&=&
		\frac{\ri}{2(\cN+1)} \Big( \frac{\cN+2}{\cN-1} \cD_\a^j G^{\a \bd i}{}_j + \ri \bar{\cD}^{\bd}_j S^{ij} \Big) ~.
		\eea
	\end{subequations}
	This defines a $\sU(\cN)$ superspace with vanishing super-Weyl tensor. For $\cN \leq 4$, it is the conformally flat limit of the supergeometry due to \cite{Howe}.
	
	In complete analogy with the $\cN=1$ story described above, the residual dilatation symmetry of conformal superspace is manifested in this framework as super-Weyl transformations. Specifically, the following combined dilatation and special conformal transformation, parametrised by a dimensionless real scalar superfield $\S$ = $\bar{\S}$, preserves the gauge $B_A = 0$:
	\begin{align}
		\mathscr{K}(\S) = \S \mathbb{D} + \hf \nabla_B \S K^{B} \quad \implies \quad B'_A = 0~.
	\end{align}
	At the $\sU(\cN)$ superspace level, this induces the following super-Weyl transformations
	\begin{subequations}
		\label{6.17}
		\bea
		\cD^{' i}_\a&=&\re^{\hf \S} \Big( \cD_\a^i+2\cD^{\b i}\S M_{\a \b} + 2 \cD_{\a}^j \S \mathbb{J}^{i}{}_j 
		- \frac{1}{2} \cD_\a^i\S {\mathbb Y} \Big)
		~,
		\label{Finite_D}
		\\
		\bar{\cD}_{i}^{' \ad}&=&\re^{\hf \S} \Big( \bar{\cD}^\ad_i+2 \bar{\cD}_{i}^{\bd} \S \bar{M}_{\ad \bd} - 2 \bar{\cD}^{\ad}_j \S \mathbb{J}^{j}{}_i 
		+ \frac{1}{2} \bar{\cD}_{i}^{\ad} \S {\mathbb Y} \Big)
		~,
		\label{Finite_DB}\\
		{\cD}'_\aa&=&\re^{\S} \Big( \cD_{\a \ad} + {\rm i} \cD^i_{\a} \S \cDB_{\ad i} + {\rm i} \cDB_{\ad i} \S \cD_{\a}^i + \Big( \cD^\b{}_\ad \S - \ri \cD^{\b i} \S \cDB_{\ad i} \S \Big) M_{\a \b} \non \\ && + \Big( \cD_{\a}{}^\bd \S + \ri \cD_{\a}^i \S \cDB^{\bd}_i \S \Big) { \bar M}_{\ad \bd} 
		- 2\ri \cD_\a^i \S \cDB_{\ad j} \S \mathbb{J}^j{}_i
		+ \frac{\ri}{8} \cD_{\a}^i \S \cDB_{\ad i} \S \mathbb{Y} 
		\Big)
		~,
		\label{Finite_DBB}
		\\
		S^{' ij}&=& \re^{\S} \Big( S^{ij}
		-\hf \cD^{ij} \S + \cD^{\a (i} \S \cD_\a^{j)} \S \Big)
		\label{Finite_S}~,
		\\
		Y^{' ij}_{\a\b}&=& \re^{\S} \Big( Y_{\a\b}^{ij}
		-\hf \cD_\a^{[i} \cD_\b^{j]} \S - \cD_\a^{[i} \S \cD_\b^{j]} \S \Big)
		\label{Finite_Y}~,
		\\
		G'_{\a\ad}&=&
		\re^{\S} \Big( G_{\a\ad}
		-{\frac{1}{4\cN}}[\cD_\a^i,\bar{\cD}_{\ad i}]\S
		-\frac{1}{2} \cD_\a^i \S \bar{\cD}_{\ad i} \S \Big)
		~,
		\label{Finite_G}
		\\
		G'_{\a\ad}{}^{i}{}_j&=& \re^{\S} \Big( G_{\a\ad}{}^{i}{}_j
		+{\frac\ri 4} \Big( [\cD_\a^{i},\bar{\cD}_{\ad j}] - \frac{1}{\cN} \d^i_j [\cD_\a^{k},\bar{\cD}_{\ad k}] \Big) \S \Big)
		~,
		\label{Finite_Gij}
		\eea
	\end{subequations}
	where we have made the definitions:
	\begin{align}
		\cD^{ij} = \cD^{\a (i} \cD_{\a}^{j)} ~, \qquad \bar{\cD}_{i j} = \bar{\cD}_{\ad (i}\bar{\cD}^{\ad}_{j)}~.
	\end{align}
	In the infinitesimal case, the corresponding transformations are a special case of the ones presented in \cite{Howe}.\footnote{Recently, the super-Weyl transformations of $\cN$-extended superspace have been described within a local supertwistor formulation approach, see \cite{HL2} for more details.} Further, for $\cN=2$, these may be read off from the finite super-Weyl transformations presented in \cite{KLRT-M2}.

\subsection{Degauging (ii): $\sSU(\cN)$ superspace}

Above we have seen that degauging the conformal superspace described in section \ref{section2.1} leads to a $\sU(\cN)$ superspace with vanishing super-Weyl tensor. The latter is characterised by the property that its local structure group is $\sSL(2,\mathbb{C}) \times \sU(\cN)_R$. As was shown in \cite{KKR}, it is actually possible to further degauge this geometry by eliminating $\sU(1)_R$ symmetry. Below, we will spell out the specifics of this procedure.

\subsubsection{$\cN=1$ case}

As can be seen via inspection of the algebra \eqref{U(1)algebra}, the $\sU(1)_R$ curvature is controlled by the torsion $X_\a$, which is the chiral field strength of an Abelian vector multiplet. It turns out to describe purely gauge degrees of freedom; by employing the super-Weyl transformations \eqref{FinitesuperWeylTf}
it is possible to fix the gauge
\bea
\label{X=0}
X_\a =0~.
\eea
Hence, since the $\sU(1)_R$ curvature is now vanishing, the corresponding connection is flat and may also be gauged away
\bea
\label{F=0}
{\F}_A =0~.
\eea

In the resulting frame, the algebra of covariant derivatives takes the form
\begin{subequations} \label{GWZalgebra}
	\bea
	\{ \cD_{\a}, \cD_{\b} \} &=& -4{\bar R} M_{\a \b}~, \qquad
	\{\cDB_{\ad}, \cDB_{\bd} \} =  4R {\bar M}_{\ad \bd}~, \\
	&& {} \qquad \{ \cD_{\a} , \cDB_{\ad} \} = -2{\rm i} \cD_{\a \ad} ~, 
	\\
	\big[ \cD_{\a} , \cD_{ \b \bd } \big]
	& = &
	{\rm i}
	{\ve}_{\a \b}
	\Big({\bar R}\,\cDB_\bd + G^\g{}_\bd \cD_\g
	- \cD^\g G^\d{}_\bd  M_{\g \d}
	\Big)
	+ {\rm i} \cDB_{\bd} {\bar R}  M_{\a \b}~,\\
	\big[ {\bar \cD}_{\ad} , \cD_{\b\bd} \big]
	& = &
	- {\rm i}
	\ve_{\ad\bd}
	\Big({R}\,\cD_{\b} + G_\b{}^\gd \cDB_\gd
	- \cDB^{\gd} G_{\b}{}^{\dd}  \bar M_{\gd \dd}
	\Big) 
	- {\rm i} \cD_\b R  {\bar M}_{\ad \bd}~,
	\eea
\end{subequations}
which describes a conformally flat GWZ superspace \cite{GWZ}. This geometry is described in terms of the complex scalar $R$ and a real vector $G_a = \overline{G_a}$ subject to the Bianchi identities
	\label{GWZBIs}
	\bea
	\bar{\cD}_{\ad} R = 0 ~, \qquad \cD_{\a}R=\bar{\cD}^{\ad}G_{\a \ad} ~, \qquad
	\cD_{(\a}{}^{\gd} G_{\b ) \gd} = 0 ~.
	\eea
	
While we utilised our super-Weyl freedom to impose the gauge \eqref{X=0}, it turns out that there is a class of residual transformations preserving this frame. The corresponding parameters are of the form
\bea 
\S=\hf \big(\s +\bar \s \big) ~, \qquad \bar \cD_\ad \s =0~, \qquad \mathbb{Y} \s = 0~.
\label{2.18}
\eea
However, in order to preserve the gauge \eqref{F=0}, 
every residual super-Weyl transformation \eqref{2.18} must be accompanied by the following compensating $\sU(1)_{R}$ transformation
\be
\cD_A \longrightarrow \re^{-\frac{1}{4} (\s - \bar{\s}) \mathbb{Y}} \cD_A \re^{\frac{1}{4} (\s - \bar{\s}) \mathbb{Y}}~.
\ee
This leads to the transformations:
\begin{subequations} 
	\label{superweylGWZ}
	\bea
	\cD'_\a &=& \re^{ {\bar \s} - \hf \s} \Big(  \cD_\a + \cD^\b \s \, M_{\a \b} \Big) ~, \\
	\bar \cD'_\ad & = & \re^{  \s -  \hf {\bar \s}} \Big(\bar \cD_\ad +  \bar \cD^\bd  {\bar \s}  {\bar M}_{\ad \bd} \Big) 
	~,\\
	\cD'_{\a\ad} &=& \re^{\hf \s + \hf \bar \s} \Big( \cD_{\a\ad} 
	+\frac{\ri}{2} \bar \cD_\ad \bar \s \cD_\a + \frac{\ri}{2} \cD_\a  \s \bar \cD_\ad  
	+ \Big( \cD^\b{}_\ad \s + \frac{\ri}{2} \bar{\cD}_\ad \bar{\s} \cD^\b \s \Big) M_{\a\b} \non \\
	&& \qquad + \Big( \cD_\a{}^\bd \bar \s + \frac{\ri}{2} \cD_\a \s \bar{\cD}^\bd \bar{\s} \Big) \bar M_{\ad \bd} \Big)~, \\
	R' &=& \re^{2\s - \bar{\s}} \Big( R +\frac{1}{4} \bar{\cD}^2 \bar{\s} - \frac{1}{4} (\bar{\cD} \bar{\s})^2 \Big) ~, \\
	G_{\a\ad}'= &=& \re^{\hf \s + \hf \bar \s} \Big( G_{\a\ad} +\ri \cD_{\a\ad} ( \s - \bar \s) + \hf \cD_\a \s \bar{\cD}_\ad \bar{\s} \Big) ~.
	\eea
\end{subequations} 
For infinitesimal $\s$, these transformations 
are a special case of those of \cite{HT}, see also \cite{Siegel:SC}.

By definition, two supergeometries are conformally related if they possess frames which are related by a finite super-Weyl transformation.
Further, a supergeometry is conformally flat if it is conformally related to the flat supergeometry.    
Then, taking $\cD_{A} = D_{A}$ to be the covariant derivatives of $\N=1$ Minkowski superspace, we may read off the conformally flat supervielbein one-forms $E^{A} = (E^{a}\,, E^{\a}\,, \bar{E}_{\ad})$ in terms of the flat superspace one-forms $\cE^{A} = (\P^{a}\,, \text{d}\q^{\a}\,, \text{d}\bar{\q}_{\ad})$ as follows
\bsubeq \label{2.29}
\begin{align}
	E^{a} &= \re^{-\frac{1}{2}(\s+\bar{\s})}\P^{a}\,, \qquad \P^{a} = \text{d}x^{a} + \ri(\q \s^{a} \text{d}\bar{\q} - \text{d}\q \s^{a}\bar{\q})\,, \label{2.29a}
	\\
	E^{\a} &= \re^{\frac{1}{2}\s - \bar{\s}}\Big(\text{d}\q^{\a} + \frac{\ri}{4}\bar{D}_{\ad}\bar{\s}\P^{\ad\a}\Big)\,,
	\\
	\bar{E}_{\ad} &= \re^{\frac{1}{2}\bar{\s} - \s}\Big(\text{d}\bar{\q}_{\ad} - \frac{\ri}{4}D_{\a}\s \P_{\ad}{}^{\a}\Big)\,,
\end{align}
\esubeq
where  $\P^a $ is the Volkov-Akulov supersymmetric one-form \cite{VA,AV}.
It should be emphasised that the one-forms $E^{A} = \text{d}z^{M}E_{M}{}^{A}$ constitute the dual basis to the vector fields $E_{A} = E_{A}{}^{M}\partial_{M}$ in the sense that 
\begin{align} \label{dual basis}
	E_{M}{}^{A} E_{A}{}^{N} = \d_{M}{}^{N}\,, \qquad E_{A}{}^{M}E_{M}{}^{B} = \d_{A}{}^{B}\,.
\end{align}

\subsubsection{$\cN>1$ case} \label{section 2.3.2}

The torsion superfield $G_{\aa}{}^i{}_j$ of $\sU(\cN)$ superspace \eqref{U(N)algebra} turns out to describe purely gauge degrees of freedom. To illustrate this point, we begin by coupling the background supergeometry to some nowhere-vanishing scalar superfield $\Xi \neq 0$ of non-vanishing dimension and $\sU(1)_R$ charge which will play the role of a conformal compensator. By performing an appropriate super-Weyl and local $\sU(1)_R$ transformation
\begin{align}
	\label{CComp}
	\Xi \longrightarrow \re^{\D_\Xi \S + \ri q_{\Xi} \r} \Xi ~, \qquad \D_\Xi, q_\Xi \neq 0~,
\end{align}
it is possible to impose the gauge $\Xi = 1$. Associated with this gauge condition are several integrability conditions. In particular, by making use of eq. \eqref{U(N)algebra-b} we see that
\begin{align}
	\label{2.34}
	\big \{ \cD_\a^i , \bar{\cD}_{\ad j} \big \} 
	 \Xi = 2 q_{\Xi}\Big( \d_j^i ( \F_\aa - G_{\aa}) + G_\aa{}^{i}{}_j\Big) = 0 ~,
\end{align}
and therefore 
\begin{align}
	\F_{\aa} = G_{\aa} ~, \qquad G_{\aa}{}^i{}_j = 0~,
\end{align}
hence the vector component of the $\sU(1)_R$ connection $\F_{\aa}$ is fixed and $G_{\aa}{}^i{}_j$ vanishes in this gauge. Actually, one can restrict our choice of compensator to a chiral one,  $\cDB^\ad_i \Xi = 0$. This condition is useful as it also fixes the spinor $\sU(1)_R$ connection
\begin{align}
	\cDB^\ad_i \Xi = q_{\Xi} \bar{\F}^\ad_i =0 \quad \implies \quad \F_\a^i = 0~, \quad \bar{\F}^\ad_i =0~.
\end{align}

Since $G_{\aa}{}^i{}_j$ is $\sU(1)_R$ neutral, this means that it can be gauged away via an appropriate super-Weyl transformation \eqref{6.17}. Thus, in general backgrounds it takes the form
\begin{align}
	G_{\a\ad}{}^{i}{}_j =-{\frac\ri 4} \Big( [\cD_\a^{i},\bar{\cD}_{\ad j}] - \frac{1}{\cN} \d^i_j [\cD_\a^{k},\bar{\cD}_{\ad k}] \Big) \bold{\S} ~,
\end{align}
for some real scalar super-Weyl inert superfield $\bm{\S}$. Hence, $G_{\aa}{}^i{}_j$ describes purely gauge degrees of freedom and so we set it to zero. In this gauge, it is natural to shift $\cD_\aa$ as follows
\bea
\cD_\aa \longrightarrow \cD_\aa+\ri G_\aa {\mathbb Y}~.
\eea

At this point the algebra of spinor covariant derivatives takes the form
\begin{subequations} 
	\bea
	\{\cD_\a^i,\cD_\b^j\}&=&
	4 S^{ij}  M_{\a\b} 
	+4\ve_{\a\b} Y^{ij}_{\g\d}  M^{\g\d}  
	-4\ve_{\a \b} S^{k[i} \mathbb{J}^{j]}{}_k
	+ 8{Y}_{\a\b}^{k(i}  \mathbb{J}^{j)}{}_k~,
	\label{acr1} \\
	\{\cD_\a^i,\cDB^\bd_j\}&=&
	-2\ri\d^i_j \cD_\a{}^\bd
	+4\d^{i}_{j}G^{\d\bd}M_{\a\d}
	+4\d^{i}_{j}G_{\a\gd}\bar{M}^{\gd\bd}
	+8 G_\a{}^\bd \mathbb{J}^{i}{}_{j}~.~~~~~~~~~
	\eea
\end{subequations} 
Hence, the $\sU(1)_R$ curvature is flat, thus the connection is flat and may be gauged away. This reduces the structure group to $\sSL(2,\mathbb{C})\times \sSU(\cN)_R$.\footnote{Due to this, such a supergeometry was referred to in \cite{KKR} as `conformally flat $\sSU(\cN)$ superspace'.} This superspace is described in terms of the complex superfields $S^{ij} = S^{ji}$, $Y_{\a \b}^{ij} = Y_{\b \a}^{ij} = - Y_{\a \b}^{ji}$, and real vector superfield $\overline{G_a} = G_a$ satisfying the Bianchi identities
\begin{subequations}
	\bea
	\cD_{\a}^{(i}S^{jk)}&=&0~, \quad
	\bar{\cD}^{\ad}_i S^{jk} = -4 \d_i^{(j} \cD_\a^{k)} G^{\a \ad} ~, 
	\\
	\cD_{(\a}^{(i}Y_{\b\g)}^{j)k}&=&0~, \quad
	\cD^{\b k} Y_{\a \b}^{ij} = - \cD_\a^{[i} S^{j]k}~, \quad
	\bar{\cD}_j^\bd Y_{\a \b}^{ij} = 2 \cD_{(\a}^i G_{\b)}{}^{\bd} ~.
	\eea
\end{subequations}
In the $\cN=2$ case this algebra of covariant derivatives coincides with the one derived by Grimm \cite{Grimm}, provided one sets the super-Weyl tensor to zero.

As in the $\cN=1$ case considered above, the gauge condition $G_{\aa}{}^i{}_j = 0$ does not completely fix the super-Weyl symmetry. Specifically, it is preserved under the class of super-Weyl transformations $\S = \hf(\s +\bar{\s})$, where $\s$ is a chiral parameter $\cDB^\ad_i \s = 0$. In order for such transformations to preserve the condition $\F_A = 0$, they must be accompanied by the compensating local $\sU(1)_R$ transformation $\cD_A \longrightarrow \re^{-\frac{1}{4} (\s - \bar{\s}) \mathbb{Y}} \cD_A \re^{\frac{1}{4} (\s - \bar{\s}) \mathbb{Y}}$. As a result, the geometry of (conformally flat) $\sSU(\cN)$ superspace is preserved by the following set of super-Weyl transformations
\begin{subequations}
	\label{SU(N)sW}
	\begin{align}
		\cD_\a^{'i}&= \re^{\frac{\cN-2}{2\cN} \s + \frac 1 \cN \bar{\s}} \Big( \cD_\a^i+ \cD^{\b i}\s M_{\a \b} + \cD_{\a}^j\s \mathbb{J}^{i}{}_j \Big) ~, 
		\\ 
		\bar{\cD}_{i}^{' \ad}&=\re^{\frac{1}{\cN} \s + \frac{\cN-2}{2\cN} \bar{\s}} \Big( \bar{\cD}^\ad_i-\bar{\cD}_{ \bd i} \bar{\s} \bar{M}^{\ad \bd} - \bar{\cD}^{\ad}_j \bar{\s} \mathbb{J}^{j}{}_i \Big)~,
		\\
		\cD_\aa' &= \re^{\hf \s + \hf \bar{\s}} \Big(\cD_\aa + \frac{\rm i}{2} \cD^i_{\a} \s \cDB_{\ad i} + \frac{\rm i}{2} \cDB_{\ad i} \bar{\s} \cD_{\a}^i + \hf \Big( \cD^\b{}_\ad (\s + \bar \s ) - \frac{\ri}{2} \cD^{\b i} \s \cDB_{\ad i} \bar{\s} \Big) M_{\a \b} \non \\ & + \hf \Big( \cD_{\a}{}^\bd (\s + \bar{\s}) + \frac{\ri}{2} \cD_{\a}^i \s \cDB^{\bd}_i \bar{\s} \Big) { \bar M}_{\ad \bd} 
		- \frac \ri 2 \cD_\a^i \s \cDB_{\ad j} \bar{\s} \mathbb{J}^j{}_i \Big) ~, \\
		S^{' ij}&= \re^{\frac{\cN-2}{\cN} \s + \frac 2 \cN \bar{\s}} \Big( S^{ij}-{\frac14}\cD^{ij} \s + \frac 1 4 \cD^{\a (i} \s \cD_\a^{j)} \s \Big)~, 
		\label{super-Weyl-S} \\
		Y^{' ij}_{\a\b}&= \re^{\frac{\cN-2}{\cN} \s + \frac 2 \cN \bar{\s}} \Big( Y_{\a\b}^{ij}-{\frac14}\cD_\a^{[i} \cD_\b^{j]}\s - \frac14 \cD_\a^{[i} \s \cD_\b^{j]} \s \Big)~,
		\label{super-Weyl-Y} \\
		G_{\aa}' &=
		\re^{\frac12 \s + \frac12 \bar{\s}} \Big(
		G_{\a\ad} -{\frac{\ri}4}
		\cD_{\a \ad} (\s-\bar{\s})
		-\frac{1}{8} \cD_\a^i \s \bar{\cD}_{\ad i} \bar{\s}
		\Big)
		~.
		\label{super-Weyl-G}
	\end{align}
\end{subequations}
In the $\cN=2$ and $\cN=3$ cases, these transformations are a special case of the ones given in \cite{KLRT-M1,KLRT-M2} and \cite{KR23}, respectively.
	
As described above for the $\N=1$ case, by choosing $\cD_{A} = D_{A}\,,$ one can express the components of the conformally flat supervielbein 
$E^A = (E^{a}\,, E^{\a}_i\,, \bar{E}_{\ad}^i)$
in terms of the flat superspace one-forms. 
Specifically, we find 
\bsubeq \label{n extend cf vierbein}
\begin{align}
	E^{a} &= \re^{-\frac{1}{2}(\s + \bar{\s})} \P^{a}\,, \qquad \P^{a} = \text{d}x^{a} + \ri(\q_{i}\s^{a}\text{d}\bar{\q}^{i} - \text{d}\q_{i}\s^{a}\bar{\q}^{i})\,,
	\label{2.40a}
	\\
	E_{i}^{\a} &= \re^{-(\frac{\N-2}{2\N}\s + \frac{1}{\N}\bar{\s})}\Big(\text{d}\q_{i}^{\a} + \frac{\ri}{4}\bar{D}_{\ad i}\bar{\s} \P^{\ad\a}\Big)\,, \label{2.40b}
	\\
	\bar{E}_{\ad}^{i} &= \re^{-(\frac{1}{\N}\s + \frac{\N-2}{2\N}\bar{\s})}\Big(\text{d}\bar{\q}_{\ad}^{i} - \frac{\ri}{4}D_{\a}^{i}\s \P_{\ad}{}^{\a}\Big) \, , \label{2.40c}
\end{align}
\esubeq
with  $\P^a $ the $\cN$-extended version of the Volkov-Akulov supersymmetric one-form \eqref{2.29a}.
As in the $\N=1$ case, these one-forms constitute the dual basis to $E_{A}$, see eq. \eqref{dual basis}.

\section{$\cN$-extended AdS superspace}
\label{section3}

As was shown in \cite{KKR}, the conformally flat superspaces described above allow for AdS supergeometries. This special case is characterised by the following conditions:
	
(i) the torsion and curvature tensors are Lorentz invariant;
	
(ii) the torsion and curvature tensors are covariantly constant.

\noindent These in turn imply the following relations:
\begin{subequations} \label{torsion conditions}
	\begin{align}
		\cN=1:& \qquad G_\aa = 0~, \qquad \cD_A R = 0~. \\
		\label{6.26} \cN>1:& \qquad Y_{\a \b}^{ij} = 0 ~, \qquad G_\aa = 0~,\qquad G_{\aa}{}^{i}{}_{j} =0~, \qquad \cD_A S^{jk} = 0~.
	\end{align}
\end{subequations}
Keeping in mind these constraints, the algebra of covariant derivatives reduces to:
\begin{subequations} 
	\label{2.170}
	\bea
	\{ \cD_\a^i , \cD_\b^j \}
	&=&
	4 S^{ij}  M_{\a\b} 	- 4 \ve_{\a\b}S^{k[i}  \mathbb{J}^{j]}{}_{k}
	~,
	\\
	\{ \bar{\cD}^\ad_i , \bar{\cD}^\bd_j \}
	&=&
	- 4 \bar{S}_{ij}  \bar{M}^{\ad\bd} 	+ 4 \ve^{\ad \bd} \bar{S}_{k[i}  \mathbb{J}^{k}{}_{j]}
	~,
	\\
	\{ \cD_\a^i , \bar{\cD}^\bd_j \}
	&=&
	- 2 \ri \d_j^i\cD_\a{}^\bd
	~,
	\\
	{[} \cD_\a^i, \cD_{\bb}{]}&=& 
	- \ri \ve_{\a \b} S^{ij} \bar{\cD}_{\bd j}
	~,
	\qquad
	{[} \bar{\cD}^\ad_i, \cD_{\bb}{]}= 
	\ri \d_\bd^\ad \bar{S}_{ij} \cD_\b^j
	~, 
	\\
	\big [ \cD_\aa , \cD_\bb \big] &=& - 2|S|^2
	(\ve_{\a \b} \bar{M}_{\ad \bd} + \ve_{\ad \bd} M_{\a \b})~,\qquad |S|^2: = \frac{1}{\cN}S^{ij} \bar{S}_{ij}>0~.
	\eea
\end{subequations}
Here $|S|$ is a constant parameter of the AdS superspace.
In the $\cN=1$ case, there is no generator $ \mathbb{J}^{i}{}_{j}$,
and the identification $\bar{R} = - S$ should be used.
Both $\sU(\cN)$ and $\sSU(\cN)$ superspaces yield the AdS geometry
\eqref{2.170}. In the $\sU(\cN)$ setting, the AdS covariant derivatives include a flat $\sU(1)_R$ connection that turns out to be useful for our analysis.
	
\subsection{Solving the constraints} \label{section3.1}
	
When $\cN>1$, the constraint $\cD_A S^{jk} = 0$ implies the following integrability condition
\begin{subequations}\label{6.42}
\begin{align}	
	\d_{(k}^{[i} S^{j]m} \bar{S}_{l)m} = 0 \quad \implies \quad 
	S^{ik} \bar{S}_{jk} =  |S|^2\d^i_j
	~.
\end{align}
Using matrix notation, the properties of $\hat{S}=(S^{ij})$ can be recast in the form:
\bea
{\mathbb S}^{\rm T} = {\mathbb S}~, \qquad {\mathbb S}^\dagger {\mathbb S}={\mathbbm 1}_{\cN}~, \qquad 
{\mathbb S}:= 
|S|^{-1}
\hat{S}~.
\eea
\end{subequations}
The properties of ${\mathbb S}$ coincide with  
those given in the formulation of 
a simple lemma proved long ago by Zumino \cite{Zumino} using the theory of matrices
\cite{Gantmacher}. The lemma states that 
\bea
{\mathbb S} = U U^{\rm T} ~, \qquad U \in \sU(\cN)~.
\label{Zum}
\eea
Here the unitary matrix $U$ is defined modulo right orthogonal shifts, 
\bea 
U ~\to ~ U \cO ~, \qquad \cO \in \sO(\cN)~.
\eea
It should be pointed out that
in the $\cN=2$ case \cite{Kuzenko:2008qw}, the general solution to 
eq. \eqref{6.42} is 
${S}^{ij} = q \, {\bm S}^{ij}$, where ${\bm S}^{ij}$ obeys the reality condition
$ \overline{{\bm S}^{ij}}= {\bm S}_{ij}$  and $q $ is a constant complex parameter of unit norm, $ |q|=1$.
By applying a rigid $\sU(1)$ phase transformation to the covariant derivatives, 
$\cD_\a^i \to q^{-1/2} \cD_\a^i$,  
one can set $q=1$.

It follows from the above analysis
that, by performing a local $\sU(\cN)_R$ transformation,\footnote{This argument requires us to work within the $\sU(\cN)$ superspace setting described in section \ref{section2.2}.}
one can bring $S^{ij}$ to the form 
\begin{align}
	\label{6.43}
	S^{ij} = \d^{ij} S ~. 
\end{align}
Now, the condition $\cD_A S^{jk} = 0$ tells us the $\sSU(\cN)_R$ connection involves only the generators 
\begin{align}
	\mathcal{J}^{ij} := - 2 \d^{k[i} \mathbb{J}^{j]}{}_k = - \cJ^{ji} 
~,
\end{align}
which
act on an $\sSU(\cN)$ `quark' $\psi^k$ as follows
\begin{align}
	\cJ^{ij} \psi^k = 2 \d^{k[i} \psi^{j]}~
\end{align}
and thus leave $\d^{ij}$ invariant. In other  words, $\cJ^{ij}$
is a generator of $\sSO(\cN)$.
Hence, in such a frame the $R$-symmetry group reduces to $\sO(\cN)_R$,
and  the $\sO(\cN)$-invariant tensor $\d^{ij}$
 may be used to raise and lower indices in accordance with
\begin{align}
	\psi^i = \d^{ij} \psi_j ~, \qquad \psi_i = \d_{ij} \psi^j~.
\end{align}
The resulting algebra of covariant derivatives is:
\begin{subequations} 
	\label{AdS}
	\bea
	\{ \cD_\a^i , \cD_\b^j \}
	&=&
	4 S \d^{ij}  M_{\a\b} + 2 \ve_{\a\b} S \cJ^{ij}
	~,
	\\
	\{ \bar{\cD}^\ad_i , \bar{\cD}^\bd_j \}
	&=&
	- 4 \bar{S} \d_{ij}  \bar{M}^{\ad\bd} - 2 \ve^{\ad\bd} \bar{S}  \cJ_{ij}
	~,
	\\
	\{ \cD_\a^i , \bar{\cD}^\bd_j \}
	&=&
	- 2 \ri \d_j^i\cD_\a{}^\bd
	~,
	\\
	{[} \cD_\a^i, \cD_{\bb}{]}&=& 
	- \ri \ve_{\a \b} S \bar{\cD}_{\bd}^i
	~,
	\qquad
	{[} \bar{\cD}^\ad_i, \cD_{\bb}{]}= 
	\ri \d_\bd^\ad \bar{S} \cD_{\b j}
		~, 
		\\
		\big [ \cD_\aa , \cD_\bb \big] &=& - 2 |S|^{2} (\ve_{\a \b} \bar{M}_{\ad \bd} + \ve_{\ad \bd} M_{\a \b})~.
		\eea
\end{subequations}
This algebra coincides with the one presented in eq. (5.27) of \cite{KKR}, which was obtained from the coset construction, provided one fixes $S = -2$.

It is instructive to compare the constraints \eqref{6.42} and their general solution \eqref{Zum} with those describing $\cN$-extended AdS supergeometries in three dimensions \cite{KLT-M12}. Following  \cite{KLT-M12}, the 3D analogue of $\sU(\cN)$ superspace is 
$\sSO(\cN)$ superspace, and the analogue of $S^{ij}$ is a covariantly constant symmetric torsion tensor $S^{IJ}$,  with $I,J= 1, \dots \cN$, which is characterised by the algebraic properties:
\bea
{\hat S}^2 =S^2 {\mathbbm 1}~, \qquad
{\hat S}:= (S^{IJ} ) ={\hat S}^{\rm T} ~, \qquad
S^2 :=\frac{1}{\cN}\, \tr ({\hat S}^2) \geq0
~.
\label{alg-constr-SS}
\eea
Since the structure group is $\sSO(\cN)$, a local  SO($\cN$) transformation can be performed  to bring  $\hat S$ to the form:
\bea \label{3d s}
S^{IJ}=S\,{\rm{diag}}(\,  \overbrace{+1,\cdots,+1}^{p} \, , \overbrace{-1,\cdots,-1}^{q} \,)
~, \qquad p+q =\cN~,
\eea
which should be compared with \eqref{6.43}. In contrast to the existence of a unique $\cN$-extended AdS superspace in four dimensions, in the 3D case there is the family of $(p,q) $ AdS superspaces, with $p+q=\cN$, see \cite{KLT-M12} for more details.

\subsection{Conformally flat realisation I} \label{cf1}

Now we provide a manifestly conformally flat realisation of $\text{AdS}^{4|4\cal N} $.
Imposing  the constraint \eqref{6.43} is not suitable for this purpose. This is why we 
relax the constraint \eqref{6.43} and return to the original algebra of 
AdS covariant derivatives, eq. \eqref{2.170}.
A conformally flat realisation of $\text{AdS}^{4|4\cal N} $ 
may be obtained by making use of the super-Weyl transformations, see eq. \eqref{superweylGWZ} and \eqref{SU(N)sW}. 

In the $\sSU(\cN)$ superspace setting, a curved supergeometry is conformally flat if the gauge freedom may be used to bring the covariant derivatives to the form
\eqref{AdSBoost}.
In such a frame, the curved covariant derivatives $\cD_A$ are obtained from the flat ones $D_A$ by applying a super-Weyl transformation. In the $\cN=1$ and $\cN=2$ cases, the definition agrees with those given in \cite{HT} (see \cite{BK} for a review) and \cite{Kuzenko:2008qw}, respectively. 
The conformal flatness of the superspace AdS$^{4|4}$ was first established by 
Ivanov and Sorin \cite{IS} using the coset  construction, see also \cite{BILS}.\footnote{Within the framework of the Wess-Zumino formulation for $\cN=1$ supergravity \cite{WZ}, this conformally flat realisation of $\text{AdS}^{4|4} $ is a simple application of the  prepotential solution of the Grimm-Wess-Zumino constraints 
\cite{GWZ} given by Siegel \cite{Siegel}, see \cite{BK} for a review.}

Our goal is to determine $\s$ corresponding to  the $\cN$-extended AdS superspace. 
For the covariant derivatives \eqref{AdSBoost},
the curvature $S^{ij}$ takes the form
\begin{subequations}
	\label{AdSCurvatures}
	\begin{align}
		S^{ij }= - \frac 1 4 \re^{\frac{\cN-2}{\cN} \s + \frac 2 \cN \bar{\s}} \Big(D^{ij} \s - D^{\a i} \s D_\a^{j} \s \Big)
=  \frac 1 4 \re^{\frac{2}{\cN} [\bar \s + (\cN -1)\s ]} D^{ij} \re^{-\s}	
 ~.  \label{ads constraints 0}
	\end{align}
Imposing the conditions 
\eqref{torsion conditions} leads to the following constraints on the chiral parameter $\s$:\footnote{As outlined in section \ref{section 2.3.2}, the transformations \eqref{3.13a} - \eqref{3.13c} imply $G_{\a\ad}{}^{i}{}_{j} = 0\,.$} 
	\begin{align} \label{ads constraints 1}
		Y_{\ab}^{ij} &= 0 \quad \implies \quad D_{(\a}^{[i} D_{\b)}^{j]} \re^\s = 0~,
		\\
		G_{\a\ad} &= 0 \quad \implies \quad [D_\a^i,\bar{D}_{\ad i}] \re^{\frac{\cN}{2}(\s + \bar{\s})} = 0 \label{ads constraints 2}
		~.
	\end{align}
In the $\cN=1$ case, the constraint \eqref{ads constraints 1} is absent, and the condition of covariant constancy of $S^{ij} $ becomes
\bea
\bar R = - \frac 1 4 \re^{2 \bar \s } D^2 \re^{-\s}=\text{const}~.
\label{ads constraints 3}
\eea
\end{subequations}
If  the constraints \eqref{ads constraints 1} and \eqref{ads constraints 2} are satisfied, 
the tensor $S^{ij}$ defined by \eqref{ads constraints 0} proves to be covariantly constant. 
 
 In the $\cN=1$ case, the constraints \eqref{ads constraints 2} and 
 \eqref{ads constraints 3} were solved in section 6.5.4 of \cite{BK}. 
For $\N=2$, the constraints \eqref{ads constraints 1} and \eqref{ads constraints 2} reduce to those given in \cite{Kuzenko:2008qw}, where they were also solved. 
We will now solve them for arbitrary $\N$. 
As $\s$ is chiral, $\bar{D}_{i}^{\ad}\s = 0\,,$ it is a function of the coordinates $x_{+}^{a} = x^{a} + \ri\q_{i}\s^{a}\bar{\q}^{i}$ and $\q_{i}^{\a}\,.$
The most general Lorentz and $\sSU(\N)_R$ invariant ansatz is then given by 
\begin{align} \label{general ansatz}
	\re^{\s} = \sum_{n=0}^{\N}A_{n}^{i_{1}\ldots i_{2n}}(x_{+}^{2})\q_{i_{1}i_{2}}\ldots\q_{i_{2n-1}i_{2n}}\,, 
\end{align}
where $\q_{ij} = \q^{\a}_{i}\q_{\a j} = \q_{ji}$ and the coefficients $A_{n}^{i_{1}\ldots i_{2n}}(x_{+}^{2})$ in general furnish reducible representations of $\sSU(\N)_R \,.$ 
Inserting \eqref{general ansatz} into \eqref{ads constraints 1} yields the following 
\begin{align}
	\re^{\s} = a + bx_{+}^{2} + s^{ij}\q_{ij}\,, 
\end{align}
where $a,b$ and $s^{ij} = s^{ji}$ are constant. 
The constraint \eqref{ads constraints 2} then implies 
\bsubeq
\begin{align}
	s^{ik}\bar{s}_{kj} &= -4a\bar{b}\d^{i}_{j}\,, \qquad \bar{s}_{ij} = \overline{s^{ij}}\,, \label{c cbar}
	\\
	a\bar{b} &= \bar{a}b\,.
\end{align}
\esubeq
Since the left-hand side of \eqref{c cbar} is positive-definite, we find that $a\bar{b} < 0\,.$
Choosing the constant $a$ to be $a=1$, the above relations lead to 
\begin{align}
	b = -\frac{1}{4\N}s^{ij}\bar{s}_{ij}\,. 
\end{align}
The solution to \eqref{ads constraints 1} and \eqref{ads constraints 2} is then
\begin{align} \label{ads soln}
	\re^{\s} = 1 - \frac{1}{4\N}s^{ij}\bar{s}_{ij}x_{+}^{2} + s^{ij}\q_{ij}\,.
\end{align}

Evaluating the torsion superfield $S^{ij}$ yields 
\begin{align} \label{stereographic s}
	S^{ij} = \frac{1}{4}\re^{\frac{2(\N-1)}{\N}\s + \frac{2}{\N}\bar{\s}}D^{ij}\re^{-\s} = s^{ij} + \cO(\q)\,. 
\end{align}
We point out that 
\bea
S^{ij}(z) \bar S_{ij} (z)  \equiv \cN |S|^2 = s^{ij} \bar s_{ij}=\text{const} ~.
\eea

Finally, given the form of the conformally flat supervielbein \eqref{n extend cf vierbein}, it follows that the spacetime metric is 
\begin{align} \label{stereo metric}
	\text{d}s^{2} = \eta_{ab}E^{a}E^{b}|_{\q=0} = \frac{\eta_{ab}\text{d}x^{a}\text{d}x^{b}}{(1-\frac{x^{2}}{4\ell^{2}})^{2}}\,, 
\end{align}
with $\ell^{-2} = |S|^2$.
It should be pointed out that, while the constraints \eqref{ads constraints 1} and \eqref{ads constraints 2} appeared in \cite{KKR} for the first time, the super-Weyl parameter \eqref{ads soln} was constructed in ref. \cite{BILS} 
by making use of an alternative approach, 
although an explicit solution of the form \eqref{AdSBoost}  was not derived in \cite{BILS}.

\subsection{Conformally flat realisation II} \label{cf2}

As can be seen from \eqref{stereo metric}, the above realisation makes use of stereographic coordinates, in which the spacetime metric is manifestly invariant under the group of four-dimensional Lorentz transformations, $\sO(3,1)\,.$ 
One can also make use of Poincar\'e coordinates, in which the spacetime metric takes the form 
\begin{align} \label{pp metric}
	\text{d}s^{2} = \left(\frac{1}{sz}\right)^{2}\left(\eta_{\ha\hb}\text{d}x^{\ha}\text{d}x^{\hb} + \text{d}z^{2}\right)\,, \qquad \ha = 0\,,1\,,3\,, 
\end{align}
and is manifestly invariant under the group of three-dimensional Poincar\'e transformations, $\sIO(2,1)\,.$
The reason for this index convention will become clear below.
This coordinate system was utilised in \cite{Butter:2012jj} to derive an alternative conformally flat realisation for the $\N=1$ and $\N=2$ AdS superspaces, and we will now extend this analysis to the case of arbitrary $\N\,.$

As described above, the relations \eqref{AdSBoost} hold for a conformally flat realisation of AdS$^{4|4\N}$. 
Our point of departure from the analysis of section \ref{cf1} is in the realisation of the flat superspace covariant derivatives. 
For this purpose, it is convenient to introduce a $3+1$ splitting of the 4D vector indices as follows.\footnote{Our notation and conventions coincide with those in \cite{Butter:2012jj}.} 
We first delete the sigma-matrix with vector index $a=2$,
\bsubeq \label{3d gm}
\begin{align}
	(\s_{a})_{\a\bd} &\longrightarrow (\g_{\ha})_{\ab} = (\g_{\ha})_{\b\a} = (\id_{2}\,, \s_{1}\,, \s_{3})\,, 
	\\
	(\tilde{\s}_{a})^{\ad\b} &\longrightarrow (\g_{\ha})^{\ab} = (\g_{\ha})^{\b\a} = \ve^{\a\g}\ve^{\b\d}(\g_{\ha})_{\g\d}\,. 
\end{align}
\esubeq
Spinor indices are raised and lowered using the $\sSL(2,\mathbb{R})$ invariant tensors $\ve^{\ab}$ and $\ve_{\ab}$, satisfying
\begin{align}
	\ve^{\ab} = -\ve^{\b\a}\,, \qquad \ve^{12} = -\ve_{12} = 1\,,
\end{align}
by the rule 
\begin{align}
	\j^{\a} = \ve^{\ab}\j_{\b}\,, \qquad \j_{\a} = \ve_{\a\b}\j^{\b}\,.
\end{align}
A four-vector $V_{a}$ can then be written as 
\bsubeq
\begin{align}
	V_{\a\bd} &= V^{a}(\s_{a})_{\a\bd} \longrightarrow V_{\ab} + \ri\ve_{\ab}V_{z}\,, \qquad V_{\ab} := V_{\ha}(\g^{\ha})_{\ab}\,, 
	\\
	V^{\ad\b} &= V^{a}(\tilde{\s}_{a})^{\ad\b} \longrightarrow V^{\ab} + \ri\ve^{\ab}V_{z}\,, \qquad V^{\ab} := V_{\ha}(\g^{\ha})^{\ab}\,. 
\end{align}
\esubeq
The flat spinor covariant derivatives then take the form 
\bsubeq
\begin{align}
	D_{\a}^{i} &= \frac{\partial}{\partial\q_{i}^{\a}} + \ri(\g^{\hm})_{\ab}\bar{\q}^{\b i}\partial_{\hm} - \bar{\q}^{i}_{\a}\partial_{z}\,,
	\\
	\bar{D}_{\a i} &= - \frac{\partial}{\partial \bar{\q}^{\a i}} - \ri(\g^{\hm})_{\ab}\q_{i}^{\b}\partial_{\hm} - \q_{\a i}\partial_{z}\,,
\end{align}
and they satisfy the anticommutation relations 
\begin{align}
	\{D_{\a}^{i}\,, D_{\b}^{j}\} = \{\bar{D}_{\a i}\,, \bar{D}_{\b j}\} = 0\,, \qquad \{D_{\a}^{i}\,, \bar{D}_{\b j}\} = -2\ri\d^{i}_{j}(\g^{\hm})_{\ab}\partial_{\hm} + 2\d^{i}_{j}\ve_{\ab}\partial_{z}\,. 
\end{align}
\esubeq
Finally, let us introduce the chiral coordinate $z_{L} := z - \q_{k}^{\a}\bar{\q}^{k}_{\a}\,, ~ \bar{D}_{\a i}z_{L} = 0\,.$ 

Given the form of the spacetime metric \eqref{pp metric}, we seek a solution to the constraints  \eqref{ads constraints 1} and \eqref{ads constraints 2} that is a function of the chiral coordinates $z_{L}$ and $\q_{i}^{\a}\,.$ 
The most general ansatz is then 
\begin{align} \label{pp ansatz}
	\re^{\s} = \sum_{n=0}^{\N}A_{n}^{i_{1} i_2 \ldots i_{2n-1}i_{2n}}(z_{L})\q_{i_{1}i_{2}}\ldots\q_{i_{2n-1}i_{2n}}\,.
\end{align}
It is an instructive exercise to check that the constraints 
are solved by 
\begin{align} \label{pp soln}
	\re^{ \s} = |s| z_{L} + s^{ij}\q_{ij}\,, \qquad 
	|s|^2 := \frac{1}{\N}s^{ij}\bar{s}_{ij} >0~,
\end{align}
for a constant symmetric tensor $s^{ij}$. The most general solution to the constraints proves to be at most quadratic in $\q$'s.

Similar to the stereographic case, evaluating the torsion superfield $S^{ij}$ yields
\begin{align}
	S^{ij} = s^{ij} + \cO(\q)\,.
\end{align}

\section{Discussion}
\label{section5}

In this paper we have elucidated the relationship between the two approaches to $\N$-extended AdS superspace developed in our previous work \cite{KKR}:
(i) the embedding formalism; and (ii) the supergravity-inspired framework.  
In the supergravity-inspired framework, the AdS supergeometry is naturally realised as a superspace with local
 $\sSL(2\,,\mathbb{C})\times\sSU(\N)$ symmetry, and is determined in terms of a covariantly constant complex isotensor $S^{ij}\,.$ 
As discussed in section \ref{section3.1}, $S^{ij}$ can be brought to the form $S^{ij} = \d^{ij}S$ by making use of a local $\sU(\N)$ transformation.\footnote{Strictly speaking, this requires us to introduce a flat $\sU(1)$ connection.} 
The structure group is then reduced to $\sSL(2\,,\mathbb{C}) \times \sO(\N)$, and the resulting algebra of covariant derivatives, eq. \eqref{AdS}, coincides with that arising from the embedding formalism \cite{KKR}. 
This provides the precise correspondence between the two approaches. 

In addition to this, we have also derived two explicit realisations of $\text{AdS}^{4|4{\cal N}} $ as a conformally flat superspace, thus extending the earlier results for $\text{AdS}^{4|4} $ \cite{IS}, $\text{AdS}^{4|8} $ \cite{Kuzenko:2008qw,Butter:2012jj}, $(p,q)$ AdS superspaces in three dimensions \cite{KLT-M12,KT}, and $\cN=1$ AdS superspace in five dimensions 
$\text{AdS}^{5|8} $ \cite{Kuzenko:2008kw}. All of these explicit results in diverse dimensions were obtained using supergravity techniques.

\subsection{Conformal flatness}

The supergravity-inspired framework introduced in \cite{KKR} and expanded upon in this work provides a powerful geometric formalism to study field theory in conformally flat superspace. This approach is  based on the concept 
of $\cN$-extended conformal superspace with flat connection \cite{Kuzenko:2021pqm} and can be used to describe every conformally flat superspace. Starting from conformal superspace and then degauging to the $\sSU(\cN)$ superspace setting, a curved supergeometry is conformally flat if the gauge freedom may be used to bring the covariant derivatives to the form \eqref{AdSBoost}.
The chiral super-Weyl factor $\re^\s$ corresponding to $\text{AdS}^{4|4{\cal N}} $ must be a solution to 
the differential constraints \eqref{ads constraints 1} and \eqref{ads constraints 2}, while the remaining  covariantly constant torsion tensor $S^{ij}$ is determined by \eqref{ads constraints 0}.
We have provided two explicit solutions to the constraints, which are described in section \ref{cf1} and \ref{cf2}.

The fact that $\text{AdS}^{4|4{\cal N}} $ is conformally flat was studied in \cite{BILS} as part of their general analysis of the conformal flatness of AdS superspaces with bodies of the form $\text{AdS}_m \times S^n$. In the $\cN=1$ case, the Maurer-Cartan equations corresponding to the coset 
$\mathsf{OSp}(1|4;\mathbb{R}) / \mathsf{SL}(2, \mathbb{C}) $ were solved in \cite{BILS} to result in the supervielbein \eqref{2.29}, where $\re^\s$ is given by \eqref{ads soln}.
To establish the conformal flatness of $\text{AdS}^{4|4{\cal N}} $ for $\cN>1$, the authors of \cite{BILS} developed 
a new approach based on embedding the AdS supergroup  
$\mathsf{OSp}(\cN |4;\mathbb{R})$ into the 
$\cal N$-extended superconformal group  $\sSU(2,2|\cN)$.\footnote{A similar construction was used in section 6.5.5 of \cite{BK} to derive the Killing supervectors of $\text{AdS}^{4|4}$. }  Making use of such an embedding allowed the authors of \cite{BILS} to reconstruct the vector supervielbein $E^a$ in the conformally flat form \eqref{2.40a},
where $\re^\s$ is given by \eqref{ads soln}. In principle, this approach could also have been used to reconstruct the spinor supervielbein  \eqref{2.40b} and \eqref{2.40c}, but this was not done in \cite{BILS}.

Essential use of the superconformal group, $\sSU(2,2|\cN)$, is common for our supergravity-inspired approach and the method proposed in \cite{BILS}. Conceptually,  however,  the two approaches are quite distinct. 

The specific feature of a conformally flat frame, eq. \eqref{AdSBoost}, is that the superspace structure group is 
$\mathsf{SL}(2, \mathbb{C}) \times \mathsf{SU}({\cal N})$. 
On the other hand, within the group-theoretic setting, 
 $\text{AdS}^{4|4{\cal N}} $ is realised as the coset superspace \eqref{1.1}
 with its structure group being $\mathsf{SL}(2, \mathbb{C}) \times \mathsf{O}({\cal N})$. 
This means that the coset construction based on the use of \eqref{1.1} is not directly suitable for obtaining a conformally flat realisation in the $\cN>1$ case. 
We elaborate on this issue in appendix \ref{section4}.

\subsection{Applications to superconformal higher-spin multiplets}

In the case of $\text{AdS}^{4|8} $, the conformally flat frame described in section \ref{cf2} has been used to study the most general $\cN=2$ supersymmetric nonlinear  sigma models in AdS${}_4$\cite{Butter:2012jj}.  
Here we sketch another application to the models for superconformal gauge multiplets given first for $\cN=1$ in \cite{KMT,KP} and later generalised to arbitrary $\cN$ in \cite{KR23}.\footnote{One may also perform an analogous analysis for the $\cN=2$ superconformal gravitino multiplet proposed in \cite{HKR}, though it will not be considered here.} These models describe the dynamics of superconformal prepotentials $\U_{\a(m) \ad(n)}$, $m,n\geq0$, in general conformally flat backgrounds. Their actions are expressed in terms of so-called `linearised higher-spin super-Weyl tensors' which play the role of gauge-invariant field strengths. Taking the background to be AdS, one may perform a degauging to write their AdS-specific form. To illustrate this point, we provide them in the $\cN=1$ and $\cN=2$ cases. The former take the form
\begin{subequations}
	\begin{align}
		\hat{\mathbb{W}}_{\a(m+n+1)}&=-\frac{1}{4} (\cDB^2 - 4 R) \cD_{(\a_1}{}^{\bd_1} \dots \cD_{\a_n}{}^{\bd_n}\cD_{\a_{n+1}} \U_{\a_{n+2} \dots \a_{m+n+1}) \bd(n)}~, \\
		\check{\mathbb{W}}_{\a(m+n+1)}&=-\frac{1}{4} (\cDB^2 - 4 R) \cD_{(\a_1}{}^{\bd_1} \dots \cD_{\a_m}{}^{\bd_m}\cD_{\a_{m+1}} \bar{\U}_{\a_{m+2} \dots \a_{m+n+1}) \bd(m)}~,
	\end{align}
\end{subequations}
while the latter are given by
\begin{subequations}
	\begin{align}
		\hat{\mathbb{W}}_{\a(m+n+2)}&=\frac{1}{48} (\cDB^{ij} + 4 \bar{S}^{ij}) \cDB_{ij} \cD_{(\a_1}{}^{\bd_1} \dots \cD_{\a_n}{}^{\bd_n}\cD_{\a_{n+1}}^i \cD_{\a_{n+2} i} \U_{\a_{n+3} \dots \a_{m+n+2}) \bd(n)}~, \\
		\check{\mathbb{W}}_{\a(m+n+2)}&=\frac{1}{48} (\cDB^{ij} + 4 \bar{S}^{ij}) \cDB_{ij} \cD_{(\a_1}{}^{\bd_1} \dots \cD_{\a_m}{}^{\bd_m}\cD_{\a_{m+1}}^i \cD_{\a_{m+2} i} \bar{\U}_{\a_{m+3} \dots \a_{m+n+2}) \bd(m)}~.
	\end{align}
\end{subequations}
Now, by making use of eq. \eqref{AdSBoost} and \eqref{AdSCurvatures}, the field strengths, and thus the models, can be written in terms of the flat derivatives $D_A$ provided one appropriately replaces the AdS-specific prepotentials $\U_{\a(m) \ad(n)}$ with their flat analogues $\bm{\U}_{\a(m) \ad(n)}$. They are related as follows:
\begin{align}
		\U_{\a(m) \ad(n)} = \re^{\frac{1}{2}(2-m-2\cN) \s + \frac{1}{2} (2-n-2\cN) \bar{\s}} \bm{\U}_{\a(m) \ad(n)}~.
\end{align}

Further applications of our approach to field theory in conformally flat backgrounds, which highlight the implications of conformal flatness, are described in appendix
\ref{AppendixC}.

\subsection{New superparticle models in $\text{AdS}^{4|4{\cal N}} $} \label{section 4.3}

The AdS supersymmetric interval $\eta_{ab} E^a E^b$
has a two-parameter deformation 
\bea \label{interval def}
\rd s^2 =
\eta_{ab} E^a E^b + \frac{1}{|S|^2} \Big( \o \ve_{\a \b} S^{ij} E^\a_i E^\b_j 
+ \bar \o \ve^{\ad \bd} \bar S_{ij} \bar E_\ad^i \bar E_\bd^j \Big)~,
\eea
with $\o$ a dimensionless complex parameter. 
This can be written as 
\begin{align}
	\rd s^2 = 
	E^{A}\eta_{AB}E^{B}\,, 
\end{align}
where the supermatrix $\eta_{AB}$ is defined as 
\begin{align} \label{supermetric def}
	\eta_{AB} = \left(
	\begin{array}{c||c|c}
		~\eta_{ab}~ & 0 & 0 \\
		\hline \hline 
		0 & \frac{\o}{|S|^{2}}\ve_{\ab}S^{ij} & 0 \\
		\hline 
		0 & 0 & \frac{\bar{\o}}{|S|^{2}}\ve^{\ad\bd}\bar{S}_{ij}
	\end{array}
	\right)\,,
	\qquad \text{Ber}(\eta) = -\left(\frac{|S|^{2}}{|\o|^{2}}\right)^{\N}\,.
\end{align}
The existence of such a supermetric means that there is a 
superparticle model
\begin{align} \label{ads model}
	S = \frac{1}{2}\int\text{d}\t\frak{e}^{-1}\left\{
	\dt{E}{}^{A}
	\eta_{AB}\dt{E}{}^{B} - (\frak{e}m)^{2} \right\}\,, 
	\qquad \dt{E}{}^{A} = \frac{\rd z^{M}}{\rd\t}E_{M}{}^{A}\,,
\end{align}
where $\mathfrak{e}$ is the einbein and $m$ the mass. In a conformally flat frame, eq. \eqref{n extend cf vierbein}, we have
\begin{align}
	\dt{E}{}^{A}\eta_{AB}\dt{E}{}^{B} &= \re^{-(\s + \bar{\s})}\dt{\P}{}^{a}\dt{\P}{}^{b}\eta_{ab} 
	\notag \\
	& \quad ~  ~ ~ + \bigg ( \frac{\o S^{ij}}{|S|^{2}}\re^{-(\frac{\N-2}{\N}\s + \frac{2}{\N}\bar{\s})}\left(\dot{\q}_{ij}
	- \frac{\ri}{2}\dot{\q}_{i}^{\a}\bar{D}_{j}^{\ad}\bar{\s}\dt{\P}_{\a\ad}
	- \frac{1}{16}\bar{D}_{\ad i}\bar{\s}\bar{D}_{j}^{\ad}\bar{\s}\dt{\P}{}^{2}
	\right) + \text{c.c.} \bigg )\,, ~
\end{align}
with $\dt{\P}{}^{a} := \dot{x}^{a} + \ri(\q_{i}\s^{a}\dot{\bar{\q}}^{i} - \dot{\q}_{i}\s^{a}\bar{\q}^{i})$ and $\dot{\q}_{ij} := \dot{\q}^{\a}_{i}\dot{\q}_{\a j}\,.$
It is evident that for $\o = 0$ we recover the standard superparticle model which is discussed in the following subsection.\footnote{Note that the expression for $\text{Ber}(\eta)$ in \eqref{supermetric def} is not well-defined for $\o = 0$, see, e.g., \cite{BK} for details.} 

In the embedding formalism, the most general superparticle model quadratic in derivatives of the evolution parameter $\t$ is given by the following
\begin{align} \label{embed model}
	S = - \frac{1}{2}\int \text{d}\t \frak{e}^{-1} \left\{\a \text{Str}(\dot{\bar{X}}\dot{X}) + \b\text{Str}(\dot{X}\dot{X}) +\bar{\b} \text{Str}(\dot{\bar{X}}\dot{\bar{X}})  + (\frak{e}m)^{2} \right\}\,,
\end{align} 
where the parameters $\a \in \mathbb{R}$ and $\b \in \mathbb{C}$ are constrained as
\begin{align}
	\a - (\b + \bar{\b}) = \frac{1}{4}\,,
\end{align}
while $X$ and $\bar{X}$ are bi-supertwistors of AdS$^{4|4\N}$, see \cite{KKR} for more details on the bi-supertwistor construction. 
The parameters are constrained in such a way that the model coincides with the bosonic one when the Grassmann variables are switched off, and the choice $\b = \ri\m\,, ~ \m \in\mathbb{R}$ yields the model proposed in \cite{KKR}.\footnote{This choice was made such that the $\m$-dependent structures generate no purely bosonic contributions.}
In the north chart, the structures present in \eqref{embed model} take the form 
\bsubeq \label{str structures}
\begin{align}
	\text{Str}(\dot{\bar{X}}\dot{X}) &= -4 \re^{-(\s + \bar{\s})}\dt{\P}{}^{2}
	\,,
	\\
	\text{Str}(\dot{X}\dot{X}) 
	&= 4\re^{-(\s+\bar{\s})}\dt{\P}{}^{2}(1 - x_{-}^{2} - 2\bar{\q}^{2} + \re^{\bar{\s}-\s}x_{+}^{2})
	\notag \\
	& \quad
	+ 16\re^{-\s}\dot{\q}_{I \a}\P^{\ad \a}
	\Big(\re^{-\s}x_{+}^{a}(\q_{I}\s_{a})_{\ad}
	+ \ri\bar{\q}_{I \ad}(1+ \re^{-\s}x_{+}^{2}) \Big)
	\notag \\
	& \quad + 8\re^{-\s}\dot{\q}_{IJ}\Big(
	\d_{IJ} -4\re^{-\s}\q_{IJ} + 4\bar{\q}_{IJ}(1+\re^{-\s}x_{+}^{2})
	-8\ri\re^{-\s}x_{+}^{a}(\q_{I}\s_{a}\bar{\q}_{J})
	\Big)\,,
\end{align}
\esubeq
with $\dot{\q}_{IJ} := \dot{\q}_{I}{}^{\a}\dot{\q}_{J\a}\,.$
Making use of the expressions \eqref{str structures} and the results of appendix \ref{section4}, actions \eqref{ads model} and \eqref{embed model} can be shown to coincide to leading order in the north chart provided one fixes 
\begin{align}
	\b = \frac{\o}{4|S|^{2}}\,. 
\end{align}

\subsection{$\k$-symmetry of the superparticle} \label{section 4.4}
Any conformally flat frame for $\text{AdS}^{4|4{\cal N}} $ has
 applications to massless superparticle models, in the spirit of \cite{BILS}. 
In a  supergravity background, the model for a massless superparticle is
\begin{align} \label{general superparticle}
	S = \frac{1}{2}\int\text{d}\t \,\frak{e}^{-1}\dt{E}{}^{a}\dt{E}{}^{b}\eta_{ab}\,,
\end{align}
where $\dt{E}{}^{a}$ can be read off from \eqref{ads model}.
In a conformally flat frame the massless superparticle model takes the form 
\begin{align} \label{cflat sp}
	S = \frac{1}{2}\int \text{d}\t \,\frak{e}^{-1} \re^{-(\s + \bar{\s})}\dt{\P}{}^{a} \dt{\P}{}^{b}\eta_{ab}\,.
\end{align}
As pointed out in \cite{BILS}, such a model is classically equivalent to the massless superparticle model 
in ${\mathbb M}^{4|4{\cal N}} $ 
by a simple redefinition of the einbein $\frak{e} \rightarrow \tilde{\frak{e}} = \re^{(\s + \bar{\s})}\frak{e}\,,$ therefore this model is invariant under $\cN$-extended superconformal transformations which scale the flat-superspace interval $ \eta_{ab} \P^a \P^b $ \cite{Sohnius}.
However, retaining the conformal factor explicitly has the advantage of keeping the symmetries of the background manifest. 
In particular, making use of the realisations \eqref{ads soln} or \eqref{pp soln}, one obtains the massless superparticle in AdS$^{4|4\N}$. 

A well-known feature of massless superparticle models is the presence of $\k$-symmetry, which was introduced in \cite{kappa1}. It was generalised to the superstring in \cite{kappa2}, see, e.g., \cite{Szg} for a review and  references. 
For the flat superparticle, these local transformations take the form 
\bsubeq
\begin{align}
	\d\q_{i}^{\a} &= -\ri\bar{\k}_{\ad i}\dt{\P}{}^{\ad\a}\,, \qquad \d\bar{\q}^{i\ad} = \ri \dt{\P}{}^{\ad\a}\k^{i}_{\a}\,,
	\\
	\d x^{m} &= \left(\q_{i}^{\a}(\s^{m})_{\a\ad}\dt{\P}{}^{\ad\b}\k^{i}_{\b} + \bar{\k}_{\ad i} \dt{\P}{}^{\ad\a}(\s^{m})_{\a\bd}\bar{\q}^{i\bd}\right)\,,
	\\
	\d \mathfrak{e} &= -4\frak{e}\left(\dot{\q}_{i}^{\a}\k^{i}_{\a} + \dot{\bar{\q}}^{i}_{\ad}\bar{\k}_{i}^{\ad}\right)\,, \label{flat e trf}
\end{align}
\esubeq
where $\k_{\a}^{i} = \k_{\a}^{i}(\t)$ is Grassmann-odd. 
This symmetry can be extended to conformally flat backgrounds by deforming the transformation \eqref{flat e trf} to the following 
\begin{align}
	\d \frak{e} = -4\frak{e}\left(\dot{\q}_{i}^{\a}\k_{\a}^{i} + \dot{\bar{\q}}_{\ad}^{i}\bar{\k}_{i}^{\ad} - \frac{\ri}{4}\bar{\k}_{\ad i} \dt{\P}{}^{\ad\b}D_{\b}^{i}\s + \frac{\ri}{4}\bar{D}_{\ad i}\bar{\s}\dt{\P}{}^{\ad\b}\k_{\b}^{i}
	\right)\,.
\end{align}
As a result, the models \eqref{ads model} and \eqref{embed model} are $\k$-symmetric in the $m = \o = 0$ case. 

\noindent
{\bf Acknowledgements:}\\
We are grateful to Igor Bandos, Dmitri Sorokin and Gabriele Tartaglino-Mazzucchelli for valuable feedback.  
The work of NEK is supported by the Australian Government Research Training Program Scholarship.
The work of SMK is supported in part by the Australian Research Council, project No. DP230101629.
The work of ESNR is supported by the Brian Dunlop Physics Fellowship.

\appendix

\section{The $\cN$-extended superconformal algebra} \label{AppendixA}

In this appendix, we spell out our conventions for the $\cN$-extended superconformal algebra of Minkowski superspace \cite{HLS}, $\mathfrak{su}(2,2|\cN)$. 
Our normalisation of the generators of $\mathfrak{su}(2,2|\cN)$ is similar to \cite{FT}.
For comprehensive discussions of the superconformal transformations in superspace, see, e.g., 
\cite{Sohnius,Ferber,Park,KTh,Raptakis}.

The conformal algebra, $\mathfrak{su}(2,2)$, consists of the translation $(P_a)$, Lorentz $(M_{ab})$, special conformal $(K_a)$ and dilatation $(\mathbb{D})$ generators. Amongst themselves, they obey the algebra
\begin{subequations} 
	\label{2.17}
	\begin{align}
		&[M_{ab},M_{cd}]=2\eta_{c[a}M_{b]d}-2\eta_{d[a}M_{b]c}~, \phantom{inserting blank space inserting} \\
		&[M_{ab},P_c]=2\eta_{c[a}P_{b]}~, \qquad \qquad \qquad \qquad ~ [\mathbb{D},P_a]=P_a~,\\
		&[M_{ab},K_c]=2\eta_{c[a}K_{b]}~, \qquad \qquad \qquad \qquad [\mathbb{D},K_a]=-K_a~,\\
		&[K_a,P_b]=2\eta_{ab}\mathbb{D}+2M_{ab}~.
	\end{align}
\end{subequations}

The $R$-symmetry group $\sU(\cN)_R$ is generated by the $\sU(1)_R$ $(\mathbb{Y})$ and $\sSU(\cN)_R$ $(\mathbb{J}^i{}_j)$ generators, which commute with all elements of the conformal algebra. Amongst themselves, they obey the commutation relations
\begin{align}
	[\mathbb{J}^{i}{}_j,\mathbb{J}^{k}{}_l] = \d^i_l \mathbb{J}^k{}_j - \d^k_j \mathbb{J}^i{}_l ~.
\end{align}

The superconformal algebra is then obtained by extending the translation generator to $P_A=(P_a,Q_\a^i,\bar{Q}^\ad_i)$ and the special conformal generator to $K^A=(K^a,S^\a_i,\bar{S}_\ad^i)$. The commutation relations involving the $Q$-supersymmetry generators with the bosonic ones are:
\begin{subequations} 
	\bea
	\big[M_{ab}, Q_\g^i \big] &=& (\s_{ab})_\g{}^\d Q_\d^i ~,\quad 
	\big[M_{ab}, \bar Q^\gd_i \big] = (\tilde{\s}_{ab})^\gd{}_\dd \bar Q^\dd_i~,\\
	\big[\mathbb{D}, Q_\a^i \big] &=& \hf Q_\a^i ~, \quad
	\big[\mathbb{D}, \bar Q^\ad_i \big] = \hf \bar Q^\ad_i ~, \\
	\big[\mathbb{Y}, Q_\a^i \big] &=&  \frac{4-\cN}{\cN} Q_\a^i ~, \quad
	\big[\mathbb{Y}, \bar Q^\ad_i \big] = \frac{\cN-4}{\cN} \bar Q^\ad_i ~, \label{2.19c} \\
	\big[\mathbb{J}^i{}_j, Q_\a^k \big] &=&  - \d^k_j Q_\a^i + \frac{1}{\mathcal{N}} \d^i_j Q_\a^k ~, \quad
	\big[\mathbb{J}^i{}_j, \bar Q^\ad_k \big] = \d^i_k \bar Q^\ad_j - \frac{1}{\mathcal N} \d^i_j \bar Q^\ad_k ~,  \\
	\big[K^a, Q_\b^i \big] &=& -\ri (\s^a)_\b{}^\bd \bar{S}_\bd^i ~, \quad 
	\big[K^a, \bar{Q}^\bd_i \big] = 
	-\ri ({\s}^a)^\bd{}_\b S^\b_i ~.
	\eea
\end{subequations}
At the same time, the commutation relations involving the $S$-supersymmetry generators 
with the bosonic operators are: 
\begin{subequations}
	\bea
	\big [M_{ab} , S^\g_i \big] &=& - (\s_{ab})_\b{}^\g S^\b_i ~, \quad
	\big[M_{ab} , \bar S_\gd^i \big] = - (\ts_{ab})^\bd{}_\gd \bar S_\bd^i~, \\
	\big[\mathbb{D}, S^\a_i \big] &=& -\hf S^\a_i ~, \quad
	\big[\mathbb{D}, \bar S_\ad^i \big] = -\hf \bar S_\ad^i ~, \\
	\big[\mathbb{Y}, S^\a_i \big] &=&  \frac{\cN-4}{\cN} S^\a_i ~, \quad
	\big[\mathbb{Y}, \bar S_\ad^i \big] =  \frac{4-\cN}{\cN} \bar S_\ad^i ~,  \label{2.20c}\\
	\big[\mathbb{J}^i{}_j, S^\a_k \big] &=&  \d^i_k S^\a_j - \frac{1}{\mathcal{N}} \d^i_j S^\a_k ~, \quad
	\big[\mathbb{J}^i{}_j, \bar S_\ad^k \big] = - \d_j^k \bar S_\ad^i + \frac{1}{\mathcal N} \d^i_j \bar S_\ad^k ~,  \\
	\big[ S^\a_i , P_b \big] &=& \ri (\s_b)^\a{}_\bd \bar{Q}^\bd_i ~, \quad 
	\big[\bar{S}_\ad^i , P_b \big] = 
	\ri ({\s}_b)_\ad{}^\b Q_\b^i ~.
	\eea
\end{subequations}
Finally, the anti-commutation relations of the fermionic generators are: 
\begin{subequations}
	\bea
	\{Q_\a^i , \bar{Q}^\ad_j \} &=& - 2 \ri \d^i_j (\s^b)_\a{}^\ad P_b=- 2 \ri \d^i_j  P_\a{}^\ad~, \\
	\{ S^\a_i , \bar{S}_\ad^j \} &=& 2 \ri  \d_i^j (\s^b)^\a{}_\ad K_b=2 \ri \d_i^j  K^\a{}_\ad
	~, \\
	\{ S^\a_i , Q_\b^j \} &=& \d_i^j \d^\a_\b \Big(2 \mathbb{D} - \mathbb{Y} \Big) - 4 \d_i^j  M^\a{}_\b 
	+ 4 \d^\a_\b  \mathbb{J}^j{}_i ~, \\
	\{ \bar{S}_\ad^i , \bar{Q}^\bd_j \} &=& \d_j^i \d^\bd_\ad \Big(2 \mathbb{D} + \mathbb{Y} \Big) + 4 \d_j^i  \bar{M}_\ad{}^\bd 
	- 4 \d_\ad^\bd  \mathbb{J}^i{}_j  ~. \label{2.21d}
	\eea
\end{subequations}

We emphasise that all (anti-)commutators not listed above vanish identically. Additionally, for $\cN=4$, the $\sU(1)_R$ generator $\mathbb{Y}$ is a central charge and may be quotiented out.

\section{Coset construction}
\label{section4}

As mentioned in the introduction, the four-dimensional $\N$-extended AdS superspace can be realised as the coset superspace \eqref{1.1}.
A key role in the study of AdS$^{4|4\N}$ as a homogeneous space is played by a local coset representative, $S$, which is an injective map $S: U \rightarrow \sOSp(\N|4;\mathbb{R})$, defined for every chart
$U$ 
of the atlas on AdS$^{4|4\N}$ chosen, with the property $\p \circ S = {\rm id}_U$, where $\p$ denotes the natural (canonical)  projection 
$\p: \sOSp(\N|4;\mathbb{R}) \to \text{AdS}^{4|4\N} =\mathsf{OSp}({\cal N}|4;\mathbb{R}) /\big[ \mathsf{SL}(2, \mathbb{C}) \times \mathsf{O}({\cal N}) \big]$. 
The atlas introduced in \cite{KKR} consists of two charts.\footnote{Given a homogeneous space $G/H$ and $\p : G\to G/H$ the corresponding natural projection, 
a global cross section $S: G/H \to G$ exists if the fibre bundle $  (G,  G/H, \p, H)$ is trivial, see, e.g., \cite{Steenrod}. Global cross-sections exist, e.g., for $\mathbb{M}^d = \sIO(d-1,1)/\sO(d-1,1)$ and $\text{AdS}_d = \sO(d-1,2)/\sO(d-1,1)$.
}
Making use of the coset representative, one can introduce the left-invariant Maurer-Cartan form, 
\begin{align}
	\o = {S}^{-1}\text{d}S\,,
\end{align}
which proves to encode the geometry of AdS$^{4|4\N}$ and takes its values in the AdS superalgebra, $\frak{osp}(\N|4;\mathbb{R})\,.$ 
The AdS superalgebra consists of supertranslation generators $W_{A} = (P_{a}\,, q_{I \a}\,, \bar{q}_{I}{}^{\ad})\,,$ Lorentz generators $M_{ab}\,,$ and $\sSO(\N)$ generators $\cJ_{IJ}\,,$ and the Maurer-Cartan form can be decomposed as the sum 
\begin{align}
	\o = \bm{E} + \bm{\O}
	\,,
\end{align}
where $\bm{E} = \bm{E}^{A}W_{A}$ is the supervielbein and $\bm{\O}$ is the connection, 
see \cite{KKR} for details.\footnote{Capital Latin letters $I\,,J\,,\ldots$ denote $\sSO(\N)$ indices and are raised and lowered with $\d^{IJ}$ and $\d_{IJ}\,.$}

In \cite{KKR}, two coordinate charts were introduced which 
naturally generalise stereographic coordinates.
We will consider the geometry in the north chart, which is parametrised by chiral coordinates 
$x_{+}^{m} = x^{m} + \ri\q_{I}\s^{m}\bar{\q}_{I}$ and $\q_{I}{}^{\m}\,.$
It is useful to introduce the following notation 
\begin{align}
	\hat{\q} = (\q_{IJ})\,, \qquad \q_{IJ} = \q_{I}{}^{\m}\q_{J \m} = \q_{JI}\,, \qquad \bar{\q}_{IJ} = \bar{\q}_{I \dmu}\bar{\q}_{J}{}^{\dmu}\,, \qquad \bar{\q}_{IJ} = \overline{\q_{IJ}}\,, 
\end{align}
and the first-order operators 
\bsubeq \label{so(n) flat d}
\begin{align} 
	D_{\m I} &= \partial_{\m I} + \ri(\s^{m})_{\m\dmu}\bar{\q}_{I}{}^{\dmu}\partial_{m}\,, 
	\\
	\bar{D}_{\dmu I} &= -\bar{\partial}_{\dmu I} - \ri\q_{I}{}^{\m}(\s^{m})_{\m\dmu}\partial_{m}\,. 
\end{align}
\esubeq
The differential operators \eqref{so(n) flat d} mimic the flat superspace covariant derivatives but carry an $\sSO(\N)$ index as opposed to an $\sSU(\N)$ index. 

Making use of these definitions, for the vector component of the supervielbein we find 
\begin{align} \label{vec vierbein}
	\bm{E}^{a} = \l \bar{\l} \bm{\P}^{m} \d_{m}{}^a\,, \qquad \bm{\P}^{m} = \text{d}x^{m} + \ri(\q_{I}\s^{m}\text{d}\bar{\q}_{I} - \text{d}\q_{I}\s^{m}\bar{\q}_{I})\,, 
\end{align}
where the chiral superfield $\l$ is defined as 
\begin{align} \label{l def}
	\l = (1 - x_{+}^{2} - 2\d^{IJ}\q_{IJ})^{-\frac{1}{2}}\,, \qquad \bar{D}_{\dmu I}\l = 0\,.
\end{align}
Making use of eq. \eqref{ads soln}, we see that $\l = \re^{-\frac{1}{2}\s}$ provided one fixes $s^{ij} = -2\d^{ij}\,,$ as described in section \ref{section3}.
The spinor component of the supervielbein was provided in \cite{KKR} only for the $\N=1$ case. 
For arbitrary $\N$ it is given by the following 
\begin{align} \label{spinor vierbein}
	\bm{E}_{I}{}^{\a} &= \re^{-\frac{1}{2}\s}(U^{-1})_{IJ}
	\bigg(\text{d}\q_{K}{}^{\a}\big(\d_{JK}+\q_{J}{}^{\n}D_{\n K}\s \big) 
	+ \bm{\P}^{\dnu \a}
	(-\re^{-\s}\q_{J}{}^{\n}(x_{+})_{\n\dnu} 
	- \ri\re^{-\bar{\s}}\bar{\q}_{J\dnu})
	\bigg)~,
\end{align}
where $U = (U_{IJ}) = U^{\rm T}$ is defined as
\begin{align}
	U = \sqrt{\id_{\N} - 4\J}\,, \qquad \J = (\J_{IJ}) = (\l^{2}\q_{IJ} + \bar{\l}^{2}\bar{\q}_{IJ}) = \J^{\rm T} \,.
\end{align}

We will now compare the present supervielbein, eq. \eqref{vec vierbein} and \eqref{spinor vierbein}, with that obtained from the supergravity-inspired approach, eq. \eqref{n extend cf vierbein}.
If the coordinate system described above is conformally flat, then the two should coincide. 
While this is true for $\N=1\,,$ which was shown in our earlier work \cite{KKR}, this is not expected to be the case for $\N\geq2$ as the local $R$-symmetry groups differ, which is elaborated upon in section \ref{section5}. 
To this end, we will specialise to the $\N=2$ case.
The spinor component of the conformally flat supervielbein, which may be extracted from \eqref{n extend cf vierbein}, takes the form
\begin{align} \label{cflat n=2}
	E_{i}^{\a} = \re^{-\frac{1}{2}\bar{\s}}\Big(\text{d}\q_{i}{}^{\a} + \frac{\ri}{4}\bar{D}_{\ad i}\bar{\s} \P^{\ad\a}\Big)\,. 
\end{align}
To compare the two expressions \eqref{spinor vierbein} and \eqref{cflat n=2}, we first consider the $\text{d}\q$ terms. 
On the one-hand, the coefficient $\re^{-\frac{1}{2}\bar{\s}}$ in \eqref{cflat n=2} is antichiral.
On the other hand, expression \eqref{spinor vierbein} has non-vanishing chiral contributions of the form 
\begin{align}
	\U_{IJ}\text{d}\q_{J}{}^{\a}\,, \qquad \bar{D}_{\ad I}\U_{JK} = 0\,,
\end{align}
where $\U = (\U_{IJ})$ is given by 
\begin{align}
	\U = \l(\id_{2} - 2\l^{2}\hat{\q} - 2\l^{4}\hat{\q}^{2} - 4\l^{6}\hat{\q}^{3} - 10\l^{8}\hat{\q}^{4})\,.
\end{align}
The two expressions therefore do not coincide, hence the frame corresponding to \eqref{vec vierbein} and \eqref{spinor vierbein} is not conformally flat.

This analysis makes use of a particular local coset representative, corresponding to the north chart of AdS$^{4|8}\,.$ 
We will now show that it is independent of the choice of coset representative. 
Let us denote the north chart by $U_{N}$, and its coset representative $S_{N}$. 
Consider a generic coset representative $S'$, with corresponding coordinate chart $U'$ such that $U' \cap U_{N} = V \neq \emptyset\,.$ 
On $V$, the two coset representatives are related as follows 
\begin{align}
	S' = S_{N}h^{-1}\,, \qquad h \in \sSL(2\,,\mathbb{C}) \times \sO(2)\,.
\end{align}
The vielbein supermatrices then satisfy
\begin{align}
	\bm{E}' = h \bm{E} h^{-1}\,,
\end{align}
where $\bm{E}'$ is the vielbein supermatrix corresponding to $U'\,.$
Since the vector component of the supervielbein, \eqref{vec vierbein}, is already in conformally flat form, we restrict our attention to those coset representatives related to $S_{N}$ by an orthogonal transformation only, in which case the spinor components \eqref{spinor vierbein} are related as 
\begin{align} \label{vierbein On}
\bm{E}'_{I}{}^{\a} = R_{IJ}\bm{E}_{J}{}^{\a}\,, \qquad R^{\T}R = \id_{2}\,, \quad R = \bar{R}\,. 
\end{align}
Now let us assume that $\bm{E}'_{I}{}^{\a}$ is in conformally flat form, \eqref{cflat n=2}. 
This means that the matrix $\tilde{R}$ relating \eqref{spinor vierbein} and \eqref{cflat n=2} satisfies the conditions \eqref{vierbein On}.
However, direct comparison of the $\text{d}\q$ terms in both expressions shows that $\tilde{R}$ is necessarily complex, hence they are not related by the rule \eqref{vierbein On}.
It follows that $\bm{E}'_{I}{}^{\a}$ cannot be in conformally flat form. 

One can consider instead the following more general condition on the matrices $R$ in \eqref{vierbein On}
\begin{align}
	R^{\dag} R = \id_{2}\,,
\end{align}
as the conformally flat form of \eqref{vec vierbein} is unspoilt by such a transformation. 
It can be shown that the matrix $\tilde{R}$ satisfies 
\begin{align}
	\tilde{R}^{\dag}\tilde{R} = \id_{2} + \cO(\q)\,,
\end{align}
and hence is not unitary.

\section{Implications of conformal flatness} \label{AppendixC}

In the conformally flat frame \eqref{AdSBoost}, many important relations in supergravity drastically simplify. As an illustration, here we concentrate on the $\cN=2$ case.  
It follows from the relations \eqref{super-Weyl-S} and \eqref{super-Weyl-Y}
that 
\begin{subequations}
\bea
S^{ij}&=& - \frac 14 \re^{ \bar{\s}} \Big( D^{ij} \s -  D^{\a (i} \s D_\a^{j)} \s \Big)~, 
	\\
		Y_{\a\b}&=& - \frac 14\re^{\bar{\s}} \Big(D_\a^{i} D_{\b i}\s + D_\a^{i} \s D_{\b i} \s \Big)~.
\eea
\end{subequations}

Of special significance in $\cN=2$ supergravity is the chiral projection operator \cite{KT-M09,Muller}
\bea
\bar{\D}
&=&\frac{1}{96} \Big((\cDB^{ij}+16\bar{S}^{ij})\cDB_{ij}
-(\cDB^{\ad\bd}-16\bar{Y}^{\ad\bd})\cDB_{\ad\bd} \Big)
\non\\
&=&\frac{1}{96} \Big(\cDB_{ij}(\cDB^{ij}+16\bar{S}^{ij})
-\cDB_{\ad\bd}(\cDB^{\ad\bd}-16\bar{Y}^{\ad\bd}) \Big)~,
\label{chiral-pr}
\eea
with $\cDB^{\ad\bd}:=\cDB^{(\ad}_k\cDB^{\bd)k}$.
Its fundamental properties are the following:
\begin{subequations} 
\bea
{\bar \cD}^{\ad}_i \bar{\D} U &=&0~, \\
\int \rd^4 x \,{\rm d}^4\q\,{\rm d}^4{\bar \q}\,E\, U
&=& \int {\rm d}^4x \,{\rm d}^4 \q \, \cE \, \bar{\D} U ~.
\label{chiralproj1} 
\eea
\end{subequations}
Here $U$ is an arbitrary super-Weyl inert scalar superfield,  $E = {\rm Ber} (E_M{}^A)$, and $\cE$ is the chiral integration measure. In the conformally flat frame, it holds that 
\begin{subequations}
\bea
E&=& 1~, \qquad \cE = \re^{-2 \s} ~,\\
\bar \D &=& \re^{2\s} \bar D^4~, \qquad
\bar{D}^4
=\frac{1}{48} \bar D^{ij}\bar D_{ij}~.
\eea
\end{subequations}

Let us consider the $\cN=2$ Gauss-Bonnet topological invariant\footnote{At the component level, $S_\c$  is a combination 
of the Gauss-Bonnet and Pontryagin invariants.} \cite{BdeWKL}
\bea
S_\c= -\int \rd^4x\, \rd^4\q\, \cE\, \Big\{ W^{\a\b}W_{\a\b}
-\X\Big\} ~,
\label{topological}
\eea
where $W_{\a\b}$ is the super-Weyl tensor, and $\X$ denotes the following chiral descendant of the torsion tensors:
\bea
\X :=  \frac{1}{6}  \bar{\cD}^{ij} \bar S_{ij}+  \bar S^{ij} \bar S_{ij}
+ \bar Y_{\dalpha \dbeta} \bar Y^{\dalpha \dbeta}~, \qquad
\bar \cD^\ad_i \X=0~.
\eea   
For every conformally flat superspace, $W_{\a\b}=0$.
In the conformally flat frame, one may show that
\bea
\X = - 2 \re^{2\s} \bar D^4 \bar \s~.
\eea
The topological nature of $S_\c$ now becomes trivial 
\bea
-\hf S_\c= \int \rd^4x\, \rd^4\q\, \bar D^4 \bar \s 
= \int \rd^4 x \,{\rm d}^4\q\,{\rm d}^4{\bar \q}\, \bar \s 
= \int \rd^4 x \,{\rm d}^4{\bar \q}\, D^4 \bar \s =0~.
\eea

Finally, let us consider the  nonlocal effective action generating the super-Weyl anomaly \cite{K2020} 
 \bea
 \G&=&  -\hf (c-a)   \int {\rm d}^4x \,{\rm d}^4 \q \, \cE 
  \int {\rm d}^4x' \,{\rm d}^4 \q' \, \cE' \, W^{\a\b}(z)W_{\a\b} (z) G_{+-} (z,z') \bar \X (z')
  ~+~{\rm c.c.} \non \\
  && -\hf a \int {\rm d}^4x \,{\rm d}^4 \q \, \cE 
  \int {\rm d}^4x' \,{\rm d}^4 \q' \, \cE' \, \X (z) G_{+-} (z,z') \bar \X (z')~,
 \label{EAction}
  \eea
with $a, c$ the anomaly coefficients. The effective action involves
two scalar Green's functions $G_{+-} (z, z') $ and $G_{-+} (z, z') $ 
that are related to each other by the rule 
\bea
G_{+-} (z, z') = G_{-+} (z', z)
\eea
and obey the following conditions: \begin{enumerate}
\item   the two-point function $G_{-+} (z, z') $ is covariantly antichiral
in $z$ and chiral in $z'$, 
\bea
\cD^i_\a  G_{-+} (z, z') =0~, \qquad \bar \cD'{}_i^\ad  G_{-+} (z, z') =0~;
\eea
\item the two-point function $G_{-+} (z, z') $ satisfies the differential equation
\bea
\bar \D G_{-+} (z, z') = \d_+ (z,z')~ .
\label{Green}
\eea
\end{enumerate}
Here we have used the  chiral delta-function 
\bea
 \d_+ (z,z'):= \bar \D\Big\{  E^{-1} \d^4(x-x') 
 \d^4 (\q-\q') \d^4 (\bar \q -\bar \q') \Big\} = \d_+(z',z)~,
 \eea
which is covariantly chiral with respect to each of its arguments,
\bea
\bar \cD^\ad_i  \d_+ (z,z')=0~, \qquad \bar \cD'{}^\ad_i  \d_+ (z,z')=0~.
\eea
Its key property is 
\bea
\J(z) =  \int {\rm d}^4x' \,{\rm d}^4 \q' \, \cE' \,\d_+(z,z')\, \J(z')~,
\qquad \bar \cD^\ad_i \J=0~,
\eea
for any covariantly chiral scalar $\J$.

In the conformally flat frame, effective action \eqref{EAction} reduces to
\bea
\G= -2a \int \rd^4 x \,{\rm d}^4\q\,{\rm d}^4{\bar \q}\, \bar \s \s~,
\eea
and becomes a local functional.

\begin{footnotesize}

\end{footnotesize}

\end{document}